\documentclass[11pt,a4paper,fleqn]{article}
\pdfoutput=1
\usepackage{graphicx}
\usepackage[usenames,dvipsnames]{xcolor} 
\usepackage{amssymb,amsmath,mathtools,mathrsfs} 
\usepackage{epsfig,subfigure} 
\usepackage{booktabs,longtable} 
\usepackage{exscale,relsize} 
\usepackage[normalem]{ulem} 
\usepackage{enumerate} 
\usepackage{jcappub}
\usepackage{caption}
\usepackage{hyperref}
\usepackage{float}
\usepackage[utf8]{inputenc}
\usepackage{comment}
\usepackage{color,colortbl}

\makeindex

\begin{document}
\hypersetup{pageanchor=false} 
\title{A general theory of linear cosmological perturbations: scalar-tensor and vector-tensor theories}

\author[a,b]{Macarena Lagos,}
\author[a,c]{Tessa Baker,}
\author[a]{Pedro G. Ferreira,}
\author[a]{Johannes Noller.}
\affiliation[a]{Astrophysics, University of Oxford, DWB,\\
Keble road, Oxford OX1 3RH, UK}
\affiliation[b]{Theoretical Physics, Blackett Laboratory, Imperial College London,\\
Prince Consort Road, London SW7 2BZ, UK}
\affiliation[c]{Department of Physics, David Rittenhouse Laboratory, University of Pennsylvania, South 33rd St, Philadelphia, PA 19104, USA}

\emailAdd{m.lagos13@imperial.ac.uk}
\emailAdd{tessa.baker@physics.ox.ac.uk}
\emailAdd{p.ferreira1@physics.ox.ac.uk}
\emailAdd{noller@physics.ox.ac.uk}

\abstract {We present a method for parametrizing linear cosmological perturbations of theories of gravity, around homogeneous and isotropic backgrounds. The method is sufficiently general and systematic that it can be applied to theories with any degrees of freedom (DoFs) and arbitrary gauge symmetries. In this paper, we focus on scalar-tensor and vector-tensor theories, invariant under linear coordinate transformations. In the case of scalar-tensor theories, we use our framework to recover the simple parametrizations of linearized Horndeski and ``Beyond Horndeski'' theories, and also find higher-derivative corrections.
In the case of vector-tensor theories, we first construct the most general quadratic action for perturbations that leads to second-order equations of motion, which propagates two scalar DoFs. Then we specialize to the case in which the vector field is time-like (\`a la Einstein-Aether gravity), where the theory only propagates one scalar DoF. As a result, we identify the complete forms of the quadratic actions for perturbations, and the number of free parameters that need to be defined, to cosmologically characterize these two broad classes of theories.}



\maketitle
\section{Introduction}
\label{sec:intro}

The possibility that we might be able to constrain General Relativity (GR) on cosmological scales is one of the science drivers behind future surveys \cite{Amendola:2012ys}. In preparation, there have been a number of proposals on how to characterize deviations from GR (or to be more specific, deviations from the $\Lambda$CDM model) in as general a fashion as possible. To some extent, the idea has been to find an approach on cosmological scales analogous to that used in the weak field, non-relativistic regime, where the Parametrized Post-Newtonian (PPN) approach captures the behaviour of a wide range of theories on the scale of the Solar System or compact binaries \cite{Berti:2015itd}. Ultimately, one would like to have a similarly systematic method for describing a general swathe of the landscape of gravitational theories on cosmological scales (see \cite{Clifton:2011jh} for a review on this topic). This description must be written in terms of a finite (and preferably small) number of ``parameters'' -- really independent functions of time -- which are easy to map onto specific theories. 

The quest for a complete and efficient parametrization is ongoing, and it is useful to briefly summarize the main approaches 
that have been considered, their strengths and weaknesses. The approach most widely used until now involves phenomenological corrections to the linear perturbation equations \cite{Amin:2007wi,Bertschinger:2008zb,Pogosian:2010tj,Bean:2010zq,Dossett:2011tn}. The Newton-Poisson equation is modified to include a time- and scale-dependent Newton's constant, and a ``gravitational slip" allows the two metric scalar potentials to differ from each other. This two-parameter approach is remarkably effective, easily implemented in Einstein-Boltzmann solvers and, with a judicious choice of functional forms, can be shown to closely mimic specific extensions of GR. While it can be shown that this parametrization is the limit of {\it any} viable theory in the quasi-static regime \cite{Silvestri:2013ne} (the cosmological equivalent of the Newtonian regime), there is no systematic way of relating it to any fundamental theory on large scales. In other words, constraints on the two parameters in this approach do not unambiguously lead to information about any putative underlying theory that might be responsible for deviations from GR. 
The same strengths and weaknesses can be found in attempts to parametrize deviations from GR in terms of a perturbed cosmological fluid \cite{Kunz:2006ca}. In this case, the parameters are the equation of state, the sound speed and terms that control adiabaticity and shear. While these parameters have a clear meaning in terms of the physics of relativistic fluids, they tell us little about what the fundamental modifications to the GR field equations or to the Einstein-Hilbert action are.

There are a number of attempts at the construction of a more fundamental parametrization. Two routes have been considered: a generalization of the field equations, in what we have called the ``Parametrized Post-Friedmann" (PPF) approach \cite{Baker:2011jy,Baker:2012zs}, or a generalization of the gravitational action, of which the two main variants are the ``Effective Action" (EA) approach \cite{Battye:2012eu,Battye:2013ida} and the ``Effective Field Theory" (EFT) approach \cite{Creminelli:2008wc,Gubitosi:2012hu,Bloomfield:2012ff,Gleyzes:2013ooa,Gleyzes:2014rba}. In the PPF approach one parametrizes the most general gauge-invariant field equations, which include up to second-order derivatives of the two scalar metric potentials. When only one scalar DoF propagates, the PPF approach covers a very general class of theories; in \cite{Baker:2012zs} it was shown that scalar-tensor, Einstein-Aether and bigravity theories are all encompassed by this parametrization. Unfortunately, as a result of its generality, there are a large number of free parameters that need to be included. Furthermore, these depend on time {\it and scale} due to the lack of knowledge of the field content of the underlying theory from which the scalar DoF comes. This makes the PPF approach potentially impractical for constraining GR on large scales.

Restricting oneself to theories that can be derived from a local fundamental action, as one does in the EFT and EA approaches, simplifies any potential parametrization. The tools of EFT have been successfully applied to characterize scalar field perturbations during inflation, allowing a systematic study of non-Gaussianity arising from higher-order operators on a quasi-de Sitter background \cite{Weinberg:2008hq,Cheung:2007st}. These ideas have been imported to late-time cosmology where, even though it is not strictly an EFT approach (one is looking at coupled but, effectively, free fields with no higher-order operators) it is useful in organizing all possible terms in the action. The approach is constructed using Arnowitt-Deser-Misner (ADM) variables and the unitary gauge to build a general spatially-invariant quadratic action for cosmological perturbations in a scalar-tensor theory; one then performs a Stueckelberg transformation to make the scalar DoF explicit and recover time diffeomorphism invariance. This is an elegant approach which has already been implemented in a couple of Einstein-Boltzmann solvers \cite{EFTCAMB1, EFTCAMB2, HiCLASS}, but is restricted to scalar-tensor theories (and particular forms of Horava-Lifschitz theory). The EA approach takes a covariant point of view {\it ab-initio}, constructing an effective action with all possible covariant combinations of the metric perturbations. It is more general than the EFT approach, is systematic and has also been implemented in existing Einstein-Boltzmann solvers \cite{Ade:2015rim}.

In this paper we would like to follow the spirit of the PPF approach and construct a systematic and general parametrization procedure, but at the level of the action, instead of the equations of motion; it will be, in some sense, an {\it integrable} version of the PPF approach. With this procedure we will construct local, general, diffeomorphism-invariant quadratic actions for linear perturbations, around homogeneous and isotropic backgrounds, encompassing all possible gravitational theories with a given field content and derivative order.
We will argue that the form of the quadratic action, crucially, depends on the gauge transformation properties of any extra fields that may arise in a modified gravity theory. An important feature of this approach is that the free parameters characterizing the quadratic action, and thus the evolution of cosmological perturbations, are defined in terms of functional derivatives of an underlying, unknown, fundamental lagrangian. One can then identify where, in the general space of parameters, a particular theory resides. As a consequence, mimicking the success of PPN, it should be straightforward to translate constraints on the general set of parameters into constraints on a particular theory (for example, Jordan-Brans-Dicke theory, Einstein-Aether gravity, bigravity, etc.). 

We will use some of the tools proposed in the EFT approach and its variants; working in terms of the 3+1 decomposition and ADM variables, connecting free coefficients with properties of fundamental theories, and assuming linear diffeomorphism invariance. However, we will not restrict ourselves to scalar-tensor theories; we will not gauge fix and, therefore, will not Stueckelberg. While in the first steps of the procedure the action that we start with will seem more complex than those proposed in the EFT approach, we show that imposing the action to be diffeomorphism-invariant rapidly simplifies it to a manageable form that is equivalent to, but more general than, other formalisms. A tremendous strength of our approach is that it is completely systematic and easily generalized to any background, degrees of freedom and gauge symmetries.
\\

\noindent {\it Outline:} We will use the main body of the paper to explain our method. Notation, intermediate results, useful asides are all relegated to appendices and appropriately referred to in the text. The structure of the paper is as follows. In Section \ref{sec:method} we explain our method, give a simple introductory example and then show the application to cosmological perturbations of a general local diffeomorphism-invariant gravitational theory, around a homogeneous and isotropic background. In Section \ref{sec:GR} we apply the method to the simplest theory - a theory with a single metric. Here we will see how linearized GR can arise from a more general construction than one would have {\it a priori} thought. In Section \ref{sec:ST} we apply the method to scalar-tensor theories and show that we recover the results of \cite{Gleyzes:2014rba} and \cite{Bellini:2014fua}. In particular, our approach includes the ``Beyond Horndeski" parameter found in \cite{Gleyzes:2014rba}, and extra parameters allowing fourth spatial derivatives of the fields in their equations of motion. In Section \ref{sec:vector} we apply our method to vector-tensor theories, with at most two derivatives of the fields. Here there are two propagating scalar DoFs, neither of which transforms as a scalar perturbation of a scalar-tensor theory. We show how to construct the most general quadratic action for perturbations with this field content and, as importantly, how to implement constraints so that we end up (as advertised) with only one propagating scalar DoF. In Section \ref{sec:discussion} we review our findings and discuss how to generalize the calculations presented in this paper.
\\

\noindent {\it Code}: Along with the paper we also release two pieces of code: Firstly the {\it xIST} package, an extension of the {\it xAct} tensor algebra system \cite{xAct}, which implements a framework to investigate general scalar-tensor theories at the level of linear perturbations. Secondly, based on {\it xIST}, a Mathematica notebook we dub {\it COPPER} (COsmological Parametrized PERturbations), which reproduces the calculations carried out in Sections 3 and 4 in this paper in detail and can be straightforwardly adapted to investigate more complicated setups. The full code and documentation can be found and downloaded at \url{https://github.com/noller/xIST}.

\section{The method: Noether identities and constraints}
\label{sec:method}

In this section we explain the method for obtaining general local quadratic actions for linear cosmological perturbations of gravitational theories with a given field content and gauge symmetries. The objective of this method is to find the maximum set of free functions parametrizing the quadratic action, and thus the cosmological predictions, of different gravitational theories. One can then automatically translate observational constraints on the free functions into constraints on these theories. In this method we will be assuming a known form for the matter sector which couples to gravity. 

Before explaining the method in detail, we first summarize the three main steps. Then we illustrate the method with a simple (non-cosmological) example of an action with a 4-vector field, invariant under $U(1)$ gauge transformations. We then proceed to analyze gravitational theories composed of at least one 2-rank tensor field, or metric, and invariant under linearized diffeomorphisms. 

The main three steps of the method are the following:
\begin{itemize}
\item[1.] Choose the fields present in the theory and the gauge symmetries to be satisfied, e.g.~invariance under linear coordinate transformations. 
\item[2.] Write down an action with all possible quadratic interactions between the fields, leading to a given maximum number of derivatives of the fields in the equations of motion.
\item[3.] Find the Noether identities associated to the required gauge symmetries, and impose the resulting constraints on the quadratic action.
\end{itemize}

Before explaining in detail, and generality, these three steps, we start with a simple example to illustrate the procedure. In particular, Step 3 above should be made clearer by this. 

\subsection{Introductory example}\label{subsec:example}

\noindent {\bf Step 1}: Consider a covariant theory for a 4-vector $A^\mu$ on Minkowski space, invariant under the following gauge transformation:
\begin{eqnarray}\label{GaugeMaxwell}
A^\alpha\rightarrow A^\alpha+\partial^\alpha \varepsilon,
\end{eqnarray}
where $\varepsilon$ is an arbitrary function of space and time.
\\

\noindent {\bf Step 2}: The most general quadratic action, leading up to second derivatives of the field in its equation of motion, can be written as: 
\begin{eqnarray}
S_A=\int d^4 x\; \left[c_1 \partial_\alpha A^\beta \partial^\alpha A_\beta+c_3 \partial_\alpha A^\beta \partial_\beta A^\alpha+m^2 A^\alpha A_\alpha\right], \label{ae}
\end{eqnarray}
where $c_1$, $c_3$ and $m$ are free constant parameters. Here we have included all possible covariant quadratic contractions of the field with an unknown coefficient in front. The structure of this action is that of the Proca-Einstein-Aether theory \cite{Jacobson:2000xp}, where we have discarded a term proportional to $(\partial_\alpha A^\alpha)^2$ (known as the ``$c_2$" term in curved space) as it is equivalent to the $c_3$ term through an integration by parts.
\\

\noindent {\bf Step 3}: If the action $S_A$ is gauge-invariant under the transformation in eq.~(\ref{GaugeMaxwell}), then a variation of the action $\delta_\varepsilon S_A$, due to an infinitesimal gauge transformation of the field, must vanish. Specifically, if we make an infinitesimal variation $\delta A^{\mu}=\partial^\mu \varepsilon$, at linear order in $\varepsilon$ we obtain: 
\begin{eqnarray}\label{DeltaSMaxwell}
\delta_\varepsilon S_A &=&\int d^4 x \left[c_1 (\partial_\alpha \partial^\beta \varepsilon\partial^\alpha A_\beta+\partial_\alpha A^\beta \partial^\alpha \partial_\beta \varepsilon)+c_3 (\partial_\alpha \partial^\beta \varepsilon\partial_\beta A^\alpha+\partial_\alpha A^\beta \partial_\beta \partial^\alpha\varepsilon)\right.\nonumber \\ & & \ \ \ \ \left.+m^2 (\partial^\alpha \varepsilon A_\alpha+A^\alpha \partial_\alpha\varepsilon)\right] \nonumber \\ &= &2\int d^4 x \left[(c_1+c_3) \partial^2\partial^\beta A_\beta-m^2 \partial^\alpha A_\alpha\right]\varepsilon,
\end{eqnarray}
where the last line comes from an integration by parts. From eq.~(\ref{DeltaSMaxwell}) we obtain a condition that must be satisfied if the action is gauge-invariant. This condition corresponds to the {\it Noether identity} associated to the gauge transformation in eq.~(\ref{GaugeMaxwell}), and is given by:
\begin{equation}
(c_1+c_3) \partial^2\partial^\beta A_\beta-m^2 \partial^\alpha A_\alpha=0,
\end{equation}
where we have used the fact that $\varepsilon$ is an arbitrary parameter, and therefore the entire bracket must vanish in order to satisfy $\delta_\varepsilon S_A=0$. 
In addition, since the action must be gauge-invariant off-shell, i.e.~for any field configuration $A^\mu$, this identity must be satisfied off-shell as well. Thus, the terms with different derivatives acting on $A_\alpha$ must vanish separately, leading to two independent constraints for the parameters:
\begin{eqnarray}
c_1+c_3&=& 0, \nonumber \\
m^2&=&0.
\end{eqnarray}
From now on, the individual constraints following from the Noether identities will be called {\it Noether constraints}. In this example, these constraints reduce the action in eq. (\ref{ae}) to that of classical electromagnetism (for an appropriate choice of normalization), which is then the most general quadratic action invariant under eq.~(\ref{GaugeMaxwell}) for a vector field with second derivatives in its equation of motion. We have systematically constructed this action by using the Noether identities to find a set of constraints on the coefficients of the original general quadratic action in eq.~(\ref{ae}).

\subsection{Gravitational action}

We will now use this method to construct the most general, linearly diffeomorphism-invariant and local quadratic action for linear perturbations of gravitational theories on a cosmological background. As already seen in the previous example, the result depends strongly on the field content and the number of their derivatives. In other words, we will be parametrizing gravitational theories with the same fields, derivative order and gauge symmetries. In general, theories that deviate from General Relativity have extra degrees of freedom, either explicitly or emerging from higher-derivative operators, extra dimensions, etc. If our method is to encompass these theories, we need to account for extra degrees of freedom.

In order to be concrete, we will sometimes refer to scalar-tensor theories to explain our procedure, but we emphasize that the method is easily generalizable to other gravitational theories. In fact, in Section \ref{sec:vector} we apply the procedure to vector-tensor theories. We follow the same three steps as above.

\vspace{3mm}
\noindent
{\bf Step 1}: Consider a gravitational theory composed of one rank-2 tensor field (or metric) and possibly some additional fields. We focus on linear perturbations around a homogeneous and isotropic cosmological background.  

The tensor degrees of freedom arise from the following perturbed metric:
\begin{equation}
g_{\alpha\beta}={\bar g}_{\alpha\beta}+\delta g_{\alpha\beta},
\end{equation} 
where ${\bar g}_{\alpha\beta}$ describes the background metric, assumed to be a spatially-flat FRW metric with a line element given by:
\begin{equation}
d \bar{s}^2=-dt^2+a(t)^2 d\vec{x}^2,
\end{equation}
where $a(t)$ is the scale factor. $\delta g_{\alpha\beta}$ describes small first-order perturbations around the background. For all the additional fields, we assume the same linearly perturbed form, with a background solution satisfying the same symmetries as $\bar{g}_{\mu\nu}$ (isotropy and homogeneity, in this case). In the case of scalar-tensor theories, with an extra scalar field $\chi$, we have
\begin{equation}
\chi = \chi_0+ \delta \chi,
\end{equation}
where $\chi_0(t)$ is the background solution of the scalar field $\chi$, and $\delta \chi$ its first-order perturbation.

We will be looking for actions which are quadratic in these perturbations and invariant under linear general coordinate transformations of the form $x^\mu\rightarrow x^\mu+\epsilon^\mu$, where $\epsilon^\mu$ is a first-order arbitrary perturbation to the coordinates $x^\mu$. Under linear coordinate transformations the background stays the same, while the linear perturbations of the metric $\delta g_{\alpha\beta}$ transform as
\begin{equation}
\delta g_{\mu\nu}\rightarrow \delta g_{\mu\nu} - {\bar g}_{\mu\beta}\partial_\nu\epsilon^\beta - {\bar g}_{\beta\nu}\partial_\mu\epsilon^\beta+\epsilon^\alpha {\bar g}_{\mu\beta}{\bar g}_{\nu\gamma} \left(\partial_\alpha {\bar g}^{\beta\gamma}\right). \label{gauge:metric}
\end{equation}
The derivation of this transformation of the metric can be found in Appendix \ref{app:gauge}. For the scalar perturbation $\delta\chi$, the corresponding transformation is
\begin{eqnarray}\label{gauge:scalar}
\delta\chi\rightarrow \delta\chi-\dot{\chi}_0\pi,
\end{eqnarray}
where $\pi$ is an arbitrary parameter corresponding to the time component of the parameter $\epsilon^\mu$, and the dot denotes a derivative with regards to the physical time $t$. 

In addition, we couple the gravitational action to matter fields. In this paper, for simplicity, we consider the matter sector to be comprised of a scalar field $\varphi$ minimally coupled to the metric, with the same gauge transformation rule as the scalar field $\chi$ in eq.~(\ref{gauge:scalar}). However, all the results found in this paper will also hold for a general perfect fluid. The formalism can also be extended to non-minimally coupled matter (for an attempt at doing this in the context of PPF see \cite{Skordis:2015yra} and in the context of EFT see \cite{Gleyzes:2015pma}).

\vspace{3mm}
\noindent
{\bf Step 2}: In this step we construct the most general local quadratic action for all the gravitational perturbation fields $\delta g_{\mu\nu}$, and any other extra field present. This quadratic action will lead to equations of motion which are linear in the perturbation fields. 

We start by assuming the existence of an underlying non-perturbative, fundamental, gravitational action $S_G$, that leads to the quadratic action we are interested in. In this paper we work in the 3+1 ADM formalism -- see Appendix \ref{app:ADM} for notation. We do this for three main reasons: i) in a cosmological setting there is a straightforward 3+1 split; ii)  for ease of comparison with EFT approaches in which time diffeomorphism invariance is broken; iii) it is straightforward to construct terms with different numbers of maximum derivatives for time and space. We emphasize, though that the procedure presented here could also be used without the ADM formalism (in a ``fully covariant" approach), although we would be forced to consider the same number of time and space derivatives. For a similar (but not identical) approach with explicit 4-dimensional covariance see \cite{Battye:2013ida}.

 In the ADM formalism we have that the metric $g_{\mu\nu}$ can be decomposed into a
lapse function $N$, shift functions $N^i$ and a 3-dimensional spatial metric $h_{ij}$ in the following way:
\begin{equation}
g_{00}=-{N^2}+h_{ij}N^iN^j, \quad g_{0i}=h_{ij}N^j, \quad g_{ij}=h_{ij}. 
\end{equation}
The underlying fundamental action will be a local functional of $N$, $N^i$, $h_{ij}$ and the extra fields, as well as their multiple time and spatial derivatives:
\begin{eqnarray}\label{SGgeneral}
S_G&=&\int d^4 x \; N\sqrt{|h|}\; L_G\left[N,N^i,h_{ij},K^i_{\phantom{i}j}, R^i_{\phantom{i}j}, \chi,\cdots\right], \label{LG1}
\end{eqnarray}
where $L_G$ is a Lagrangian functional, $|h|$ is the determinant of $h_{ij}$, and the ellipses encompass higher derivatives of the metric and any extra field(s). Given that we are seeking a linearly diffeomorphism-invariant action, we have replaced time derivatives and secondary spatial derivatives of the 3-dimensional metric in $L_G$ by the extrinsic curvature tensor $K^i_{\phantom{i}j}$ and the intrinsic 3-dimensional curvature $R^i_{\phantom{i}j}$, respectively. In general, we will consider $S_G$ to be a functional of a set of building blocks $\vec{\Theta}=(N, N^i, h_{ij}, K^{i}_{\phantom{i}j}, R^{i}_{\phantom{i}j}, \chi, \cdots)$. It is important to note that, since the building blocks include all time and spatial derivatives of the fields, we have to make an extra assumption on $S_G$, otherwise we could have infinitely many of these terms. We will choose a maximum number of (combined space and time) derivatives allowed for the fields in the equations of motion (and thus in the action) and truncate at that order. 

Given that we are interested in linear perturbations of the gravitational fields, we need the quadratic expansion of $S_G$ in $\delta N$, $\delta N^i$, $\delta h_{ij}$, $\delta\chi$ and the rest of the extra fields. To do so, we take the functional Taylor expansion of $L_G$ around the background fields in terms of the perturbed set of building blocks $\delta {\vec \Theta}= (\delta N,\delta N^i,\delta h_{ij},\delta K^i_{\phantom{i}j}, \delta R^i_{\phantom{i}j}, \delta\chi, \cdots)$, so that:
\begin{eqnarray}\label{ExpansionLgrav}
L_G & \simeq &{\bar L}+ L_{\Theta_A} \delta\Theta_A +\frac{1}{2} L_{ \Theta_A\Theta_B}\,
\delta\Theta_A \delta\Theta_B
\end{eqnarray}
where $\bar{L}$ is the zeroth order Lagrangian ($L_G$ evaluated at the background), and the subindices $A$ and $B$ label the different building blocks. The terms $L_{\Theta_A}$ and $ L_{ \Theta_A\Theta_B}$ are what we call {\it coefficients}, and are given by functional derivatives of $L_G$ evaluated at the background; therefore they generally depend on time. Explicitly, $L_{\Theta_A}\equiv \partial L_G/\partial \Theta_A$ and $ L_{\Theta_A\Theta_B}\equiv \partial^2 L_G/\partial \Theta_A\partial \Theta_B$. Notice that even though the fields ($g_{\mu\nu}$, $\chi$, etc) have only linear perturbations, the perturbed building blocks could have higher-order perturbations as result. Thus, we clarify that $\delta \vec{\Theta}$ contains both first and second-order perturbative pieces.

We can now find the Taylor expansion of the gravitational action, which is given by:
\begin{align}\label{ExpansionLG}
S_G& \simeq \int d^4x\; \left[ a^3{\bar L}+\delta_1\left(N\sqrt{|h|}\right){\bar L}+a^3 L_{\Theta_A} \delta\Theta_A+\delta_2\left(N\sqrt{|h|}\right){\bar L}\right.\nonumber \\
& \left. + \delta_1\left(N\sqrt{|h|}\right) L_{\Theta_A} \delta\Theta_A+\frac{1}{2}a^3 L_{ \Theta_A\Theta_B} \delta\Theta_A\delta\Theta_B\right],
\end{align}
where $\delta_n$ stands for an n$^{\rm th}$ order perturbation. In addition, we include a matter action: 
\begin{eqnarray}
S_M&=&\int d^4 x \; N\sqrt{|h|}\; L_M\left[N,N^i,h_{ij},\varphi,\cdots\right],
\end{eqnarray}
where $L_M$ is once again a Lagrangian functional. Here, we have generically represented ``matter fields" with $\varphi$, but they can be fields of any spin, perfect or imperfect fluids, etc. This action is assumed to be known, and therefore its Taylor expansion can be carried out straightforwardly. The linear terms of the Taylor expansion of the total action $S_G+S_M$ will be zero, and will lead to the background equations of motion (see Appendix \ref{App:Background}), while the quadratic terms will give the total quadratic action $S_G^{(2)}+ S_M^{(2)}$ determining the evolution of the cosmological perturbations. Explicitly, the second-order gravitational action will be given by:
\begin{equation}\label{QuadraticLG}
S_G^{(2)}=\int d^4x\; \left[a^3 L_{\Theta_A} \delta_2\Theta_A+\delta_2\left(N\sqrt{|h|}\right){\bar L}+
\delta_1\left(N\sqrt{|h|}\right) L_{\Theta_A} \delta_1\Theta_A+\frac{1}{2}a^3 L_{ \Theta_A\Theta_B}
\delta_1\Theta_A\delta_1\Theta_B\right],
\end{equation}
where we have used eq.~(\ref{ExpansionLG}) and the fact that a given perturbed building block can be separated into a first and second-order perturbation as: $\delta \Theta_A= \delta_1 \Theta_A+\delta_2 \Theta_A$. As we will see in the next section, only some building blocks $ \Theta_A$ will have a second-order perturbation, (for example, the lapse $N$ or the 3-curvature, $R$). As we have already mentioned, the coefficients $L_{\Theta_A}$ and $ L_{ \Theta_A\Theta_B}$ can be derived from the fundamental non-perturbative action. However, we will assume that such an action is not known and thus these coefficients will be left as free functions to be fixed by the Noether constraints, in a way analogous to the coefficients $c_1$, $c_3$ and $m^2$ in the example presented in Section \ref{subsec:example}.

\vspace{3mm}
\noindent
{\bf Step 3}: In this step we impose that the total quadratic action (from gravity and matter) is invariant under linear coordinate transformations. We do so by finding the relevant Noether identities, and solving the resulting Noether constraints. 

To find the Noether identities, we write down all the perturbed building blocks $\delta \Theta_A$ in terms of the perturbation fields $\delta g_{\mu\nu}$, $\delta\chi$, etc.~and vary the quadratic action with regards to them. Specifically, in this paper, we vary the quadratic action in terms of the scalar-type perturbation fields, according to the standard Scalar-Vector-Tensor (SVT) decomposition of fields \cite{1992PhR...215..203M}. We focus only on these scalar perturbations, as they are the seeds of large-scale structure in the density field, and therefore cosmologically relevant. We can ignore the vector and tensor perturbations as they decouple from the scalar perturbations on a homogeneous and isotropic background. Thus, we write $\delta g_{\mu\nu}$ as:
\begin{equation}
\delta g_{00}=-2\Phi, \; \delta g_{0i}= \partial_i B, \; \delta g_{ij}=a^2\left[-2\Psi \delta_{ij}+2\partial_i\partial_j E\right],
\end{equation}
where we have four scalar perturbation fields $\Phi$, $B$, $\Psi$ and $E$. From eq.~(\ref{gauge:metric}), we can find how these scalars will transform under linear coordinate transformations:
\begin{eqnarray}\label{gauge:scalarsmetric}
\Phi&\rightarrow&\Phi-{\dot \pi}, \nonumber\\
\Psi&\rightarrow&\Psi +H\pi, \nonumber\\
B&\rightarrow&B +\pi-a^2{\dot \epsilon}, \nonumber\\
E&\rightarrow&E -\epsilon,
\end{eqnarray}
where $H=\dot{a}/a$. Here we have used the SVT decomposition on the 4-vector $\epsilon^\mu$ to rewrite it as: $\epsilon^\mu=(\pi,\epsilon^{Ti}+\delta^{ij}\partial_j\epsilon)$, such that $\partial_i \epsilon^{Ti}=0$. We notice that this 4-vector has two scalar-type parameters $\pi$ and $\epsilon$ that only affect the scalar-type perturbations of the quadratic action, and one vector-type parameter $\epsilon^{Ti}$ that will not be relevant for our calculations. With this decomposition in hand, we can rewrite all the perturbed building blocks depending on the metric in terms of these four scalar perturbations (see Appendix \ref{app:metricperts} for a full list), and obtain a quadratic action $S_G^{(2)}+S_M^{(2)}$ in terms of $\Phi$, $B$, $\Psi$, $E$, $\delta \varphi$, and the rest of the perturbed extra fields. 

We now take an infinitesimal variation of the total quadratic action with regards to each one of the scalar-type perturbation fields. For scalar-tensor theories, where the matter sector is comprised by a scalar field $\varphi$, the variation of the quadratic action can be written as:
\begin{eqnarray}
\delta S_G^{(2)}+\delta S_M^{(2)}=\int d^4 x\; \left[{\cal E}_\Phi\delta \Phi+{\cal E}_B\delta B+{\cal E}_\Psi\delta\Psi+{\cal E}_E\delta E+ {\cal E}_\chi\delta \chi + {\cal E}_\varphi\delta \varphi \right] \label{epsilonscalars}
\end{eqnarray}
where ${\cal E}_X$ is the equation of motion for the perturbation field $X$. 
To find the Noether identities, we replace the variations of the fields by the corresponding gauge transformations in eq.~(\ref{gauge:scalarsmetric}) and (\ref{gauge:scalar}), and integrate by parts to end up with: 
\begin{eqnarray}
\delta_{g} S_G^{(2)}+\delta_{g} S_M^{(2)} &=&\int d^4x\; \left[{\cal E}_B+H{\cal E}_\Psi+{\dot {\cal E}}_\Phi-{\cal E}_\chi \dot{\chi}_0- {\cal E}_\varphi \dot{\varphi}_0\right]\pi \\
 &+&\int d^4x\; \left[-{\cal E}_E+\frac{d}{dt}\left(a^2{\cal E}_B\right)\right]\epsilon,
\end{eqnarray}
where the expression $\delta_g$ stands for a variation of the action due to a gauge transformation. We have used the fact that the matter perturbation field $\delta \varphi$ transforms in an analogous way to $\delta \chi$. Given that the total quadratic action is invariant under these gauge transformations, and given that both $\pi$ and $\epsilon$ are arbitrary and independent, each set of brackets must be zero; this gives us the two \textit{Noether identities} associated to the two scalar gauge parameters of the model $\pi$ and $\epsilon$. Furthermore, each combination of coefficients, inside each of the brackets, multiplying the perturbation fields and their derivatives such as $\Phi$, $ {\dot \Phi}$, $\partial^2 \Phi$, $\Psi$, etc, must be individually zero for the Noether identities to be satisfied off-shell, giving a set of \textit{Noether constraints}.
These constraints will be, in general, linear ordinary differential equations of the coefficients $L_{\Theta_A}$ and $ L_{ \Theta_A\Theta_B}$. However, for all the cases presented in this paper, these Noether constraints can be solved algebraically. Solving all of these constraints and replacing the solutions in the quadratic action allows us to determine the number of independent free coefficients and the number of degrees of freedom of the theory. The resulting action will be the most general linearly diffeomorphism-invariant local quadratic action, given the field content. It is important to remark that even though we only impose gauge invariance under the scalar gauge parameters $\pi$ and $\epsilon$, the resulting action will also be gauge-invariant under the vector gauge parameter $\epsilon^{Ti}$, and thus it will be fully invariant under linear coordinate transformations. In turn, this means that the independent free parameters characterizing the action for scalar perturbations are the same as those characterizing the vector and tensor perturbations.
\vspace{3mm}
 
We emphasize that the procedure described above is easily generalizable to different backgrounds, to include extra gravitational fields, and different gauge symmetries. To illustrate this, in the next sections we apply the procedure to gravitational actions including a metric and one extra scalar field or vector field. We will also briefly discuss a case in which we impose an extra gauge symmetry, as well as linear diffeomorphism invariance, in the quadratic action. 

As a comparison, we mention that in the EFT approach, the quadratic action for scalar-tensor theories is constructed by working in the unitary gauge, which simplifies calculations, as the dependence on the scalar field vanishes and thus the action only depends on the metric. In this situation one constructs a spatially gauge-invariant quadratic action, and in the end gauge transforms (or ``Stueckelberg'') to make the extra scalar field explicit, and recover the time gauge invariance \cite{Gubitosi:2012hu,Bloomfield:2012ff,Gleyzes:2013ooa}. However, a generalization of this procedure is not straightforward. For example, in general, in a vector-tensor theory we would have to fix the spatial and time gauge invariance in order to eliminate the entire dependence on the vector field, and have a quadratic action depending only on the metric. It is not clear that the construction of such metric action is simple as now there would be no gauge symmetry satisfied, relating the different coefficients of the action. In addition, in general, in bimetric theories there is no way of using the gauge freedom to eliminate the entire dependence on the second metric field.

Returning to our procedure, it is possible to easily generalize the matter content to encompass fluids such as baryons, dark matter, etc. In such cases it is convenient to work at the level of the equations of motion instead of the quadratic action. We can do this by finding the first-order equations of motion $\mathcal{E}^{\mu\nu}$: 
\begin{equation}
\mathcal{E}^{\mu\nu}\equiv \frac{\delta S_G^{(2)} }{\delta g_{\mu\nu}} = -\frac{\delta_1(\sqrt{-g})}{2}{\bar T}^{\mu\nu}
- \frac{a^3}{2}\delta_1 T^{\mu\nu},
\end{equation}
where we have expanded the energy-momentum tensor of the matter content up to first order, $T^{\mu\nu}={\bar T}^{\mu\nu}+ \delta_1 T^{\mu\nu}$. Note that for finding $\mathcal{E}^{\mu\nu}$ we make a variation of the quadratic gravitational action $S_G^{(2)}$ only. We then have that the equations of motion for each one of the scalar metric perturbation fields become:
\begin{align}
\mathcal{E}_\Phi &= \left(\delta_1\sqrt{-g}\right){\bar T}^{00} + a^3\delta_1 T^{00}, \nonumber \\
\mathcal{E}_B &=\partial_i\left(\delta_1\sqrt{-g}\right) {\bar T}^{0i} + a^3\partial_i\left(\delta_1 T^{0i}\right), \nonumber \\
\mathcal{E}_\Psi &= \left[\left(\delta_1\sqrt{-g}\right){\bar T}^{ij} + a^3\delta_1 T^{ij}\right]\bar{h}_{ij}, \nonumber\\
\mathcal{E}_E &= -a^2\left[\partial_i\partial_j\left(\delta_1 \sqrt{-g}\right){\bar T}^{ij} + a^3\partial_i\partial_j\left(\delta_1 T^{ij}\right)\right]. \label{fulleom}
\end{align} 
Naturally, we need to supplement the system with the equations of motion of the matter fields that constitute $T^{\mu\nu}$.

As we have mentioned before, this procedure is useful for translating cosmological constraints into constraints on fundamental gravitational actions, and for easily finding where a given gravity theory lies in the space of free parameters. However, we point out that even accurate observational constraints on the set of parameters do not lead uniquely to one fundamental theory. As we will see in the next sections, there is a considerable degeneracy of the parameters $L_{\Theta_A}$ and $L_{\Theta_A\Theta_B}$ that lead to the same observable combinations. The reason for this degeneracy is that we are only constraining the linear evolution of perturbations, but a corresponding higher-order theory could take different forms. 

In the following sections we apply the procedure presented above to the simplest (and well-established) cases of General Relativity, scalar-tensor and vector-tensor theories, as they will allow us to illustrate the method in familiar settings. In a later paper we will apply our method to a wider range of theories.

\section{Recovering General Relativity}
\label{sec:GR}
In this section we parametrize linearly diffeomorphism-invariant gravitational theories containing only one metric field, coupled minimally to a scalar field that constitutes our matter sector. However, the results presented in this section also hold for a perfect fluid matter sector. As in the previous section, we analyze linear perturbations of the fields around a homogeneous and isotropic background.

We follow Step 2 for constructing the most general quadratic gravitational action for a metric. We will allow, at most, second-order derivatives in the equations of motion for the perturbation fields. We start by writing down all the possible perturbed building blocks $\delta \Theta_A$ on which the Taylor-expanded Lagrangian $L_G$ might depend. In the ADM formalism, we have $\delta \vec{\Theta}=(\delta N,\delta {\dot N}, \delta \partial_i N, \delta \partial_i\dot{N}, \delta \partial_i\partial_j N, \delta N^i, \delta {\dot N}^i, \delta \partial_j N^i,$ $ \delta \partial_j \dot{N}^i, \delta \partial_i\partial_j N^k, \delta h_{ij}, \delta K^i_{\phantom{i}j},\delta R^i_{\phantom{i}j})$, where the latter two terms replace $\dot{h}_{ij}$ and $\partial_k\partial_l h_{ij}$. As expected, here we have included all possible metric perturbations up to two derivatives\footnote{For simplicity, we have not considered here terms with first spatial derivatives of $\delta h_{ij}$. However, they can be systematically added, and the quadratic actions given in eq.~(\ref{LT0})-(\ref{LT2}) will not change except for the explicit relations between the coefficients $T_*$ and $L_*$.}. Note that partial derivatives of the perturbation fields are taken with regards to the background metric, and thus we raise and lower the indices of the perturbed building blocks with $\bar{h}_{ij}$. Also, $\delta$ commutes with partial spatial derivatives and so, for instance, $\delta (\partial_i N) = \partial_i (\delta N)$. We emphasize that, contrary to GR, we are \textit{a priori} assuming that $\delta N$ and $\delta N^i$ could in principle be dynamical fields (with time derivatives); we will let the Noether identities dictate whether they really are or not. As we will see later, the Noether constraints will indeed make $\delta N$ and $\delta N^i$ be non-dynamical fields.

We now proceed to Taylor expand $L_G$ up to second order in the perturbed building blocks, as in eq.~(\ref{ExpansionLgrav}). A few comments are in order that will help us understand the notation in the calculations that follow. In the subscripts of the coefficients $L_{\Theta_A}$ and $L_{\Theta_A\Theta_B} $ (hereafter referred to as $L_*$), we use `$S$' (for ``Shift") as a proxy for $N^k$, and $\partial^n$ to signal the number of spatial derivatives acting on the ADM metric variables. We recall that all coefficients $L_*$ are evaluated at the level of the background and thus can only depend on $\bar{N}=1$ and $\bar{h}_{ij}$. Therefore, statistical isotropy allows us to discard coefficients with an odd number of indices (it is not possible to construct such an object out of $\bar{h}_{ij}$ and $\bar{N}$ that respects the isotropy) and imposes symmetries on coefficients with an even number of indices. We then use the following notation for the coefficients $L_*$:
\begin{align}\label{DefL2D}
 L_{A^i_{\phantom{i}j}}&=L_A\delta^j_{\phantom{i}i}  ,\quad  L_{A_iB_j}= L_{AB}{\bar h^{ij}}, \quad  L_{AB^i_{\phantom{i}j}}=L_{AB}\delta^j_{\phantom{j}i},\nonumber \\
 L_{A^i_{\phantom{i}j}B^r_{\phantom{r}s}}&=L_{AB+}\delta^j_{\phantom{j}i}\delta^s_{\phantom{s}r} +L_{AB\times}\left(\delta^j_{\phantom{j}r}\delta^s_{\phantom{s}i}+{\bar h}^{js}{\bar h}_{ir}\right), \; \mbox{where $A_{ij}=A_{ji}$ (and/or $B_{ij}=B_{ji}$)} \nonumber\\
  L_{A^i_{\phantom{i}j}B^r_{\phantom{r}s}}&=L_{AB+}\delta^j_{\phantom{j}i}\delta^s_{\phantom{s}r} +L_{AB\times 1}\delta^j_{\phantom{j}r}\delta^s_{\phantom{s}i}+L_{AB\times 2} {\bar h}^{js}{\bar h}_{ir} , \nonumber\\
 L_{B^lA^i_{\phantom{i}jk}}& =L_{BA\times 2} \bar{h}_{li}\bar{h}^{jk}+L_{BA\times 1} \left(\delta^j_{\phantom{j}l}\delta^k_{\phantom{k}i}+\delta^k_{\phantom{k}l}\delta^j_{\phantom{j}i}\right), \; \mbox{where $A^i_{\phantom{i}jk}=A^i_{\phantom{i}kj}$}, 
\end{align}
where $A$, $A^i_{\phantom{i}j}$, etc.~represent any term of the building blocks with the corresponding index structure. Two exceptional cases that do not follow the previous definitions are these:
\begin{align}\label{DefL2D2}
 L_{\partial_i\partial_j  N \partial_s N^r }&= \frac{1}{3}L_{\partial^2 N\partial S}\left( \bar{h}^{ji}\delta^s_{\phantom{s}r}+ \delta^j_{\phantom{j}r}\bar{h}^{si}+{\bar h}^{js}\delta^i_{\phantom{i}r} \right),\nonumber\\
  L_{\partial_l N \partial_j\partial_k N^i}&=\frac{1}{3}L_{\partial^2S\partial N}\left( \delta^l_{\phantom{l}i}\bar{h}^{jk}+\bar{h}^{jl}\delta^k_{\phantom{k}i}+\bar{h}^{kl}\delta^j_{\phantom{j}i}\right).
\end{align}
With all these definitions in hand we can Taylor expand $L_G$ up to second order. Recall that the perturbed building blocks can contain first- and second-order perturbations of the metric. However, we find that only $N$, $\sqrt{|h|}$, $R$ and $K$ have second-order terms; thus from now on we use $\delta$ for first-order perturbations and $\delta_2$ for second-order perturbations, unless explicitly stated otherwise.

We then can find the Taylor expansion of the gravitational action $S_G$, as in eq.~(\ref{ExpansionLG}). We require the first and second-order perturbations of the metric density:
\begin{eqnarray}
\delta_1 \left(N\sqrt{|h|}\right)&=&\delta\sqrt{|h|}+a^3\delta N, \nonumber \\
\delta_2 \left(N \sqrt{|h|}\right)&=&\delta_2\sqrt{|h|}+a^3\delta_2 N +\delta\sqrt{|h|}\delta N,
\end{eqnarray}
where we have used $\sqrt{|\bar h|}=a^3$ and $\bar{N}=1$. To simplify notation we introduce $\delta h^i_{\phantom{i}j}\equiv \bar{h}^{ik}\delta h_{kj}$ and $\delta h\equiv \bar{h}^{ij}\delta h_{ij}$, and thus $\delta$(trace of $h_{ij}$) $\neq$ (trace of $\delta h_{ij}$), which will be used later. 

Finally, the action for our matter sector scalar field $\varphi$ is given by:
\begin{equation}
S_M=-\int d^4x \sqrt{-g}\left( \frac{1}{2}\partial_\mu \varphi \partial^\mu \varphi + V(\varphi)\right),
\end{equation}
where $V(\varphi)$ is some potential. This action can be straightforwardly written in terms of the ADM variables, and Taylor expanded up to second order in the linear perturbations of the metric and scalar field.

As mentioned in the previous section, from the linear expansion of the total action (gravity and matter) we find the background equations. In this case, from the metric perturbations, we get:
\begin{eqnarray}
{\bar L}+L_N-3HL_{\dot N}-{\dot L}_{\dot N}-3HL_K&=&\rho_m, \nonumber \\
{\bar L}-3HL_K-{\dot L}_K+2L_h&=& -P_m, \label{eq:back}
\end{eqnarray}
where $\rho_m$ and $P_m$ are the energy density and pressure of the fluid, respectively. Explicitly, 
\begin{equation}
\rho_m=\frac{1}{2}\dot{\varphi}_0^2+V_0, \quad P_m=\frac{1}{2}\dot{\varphi}_0^2-V_0,
\end{equation}
where the subscript $0$ indicates the background value. Equations (\ref{eq:back}) are a generalization of the background equations shown in \cite{Gleyzes:2014rba}, whose explicit derivation can be found in Appendix \ref{App:Background}. Note that we will also have an additional background equation from the linear terms in the matter sector field:
\begin{equation}
\ddot{\varphi}_0+3H\dot{\varphi}_0+V_0'=0, \label{N4}
\end{equation} 
where $V_0'$ is the derivative of the potential with regards to the scalar field, evaluated at the background.

On the other hand, from the quadratic terms of the total action, we obtain the action that governs the evolution of the cosmological perturbations. In this case, the full quadratic gravitational action in eq.~(\ref{QuadraticLG}) can be written as:
\begin{eqnarray}
S_G^{(2)}=\int d^4 x \sum_{i=0}^2 {\cal L}_T^{i}. \label{GRaction}
\end{eqnarray}
The subscript $T$ here stands for ``tensor", as in the present case we only have a tensor field. The ${\cal L}_T^{i}$ are quadratic Lagrangians leading to $i$ derivatives of the perturbation fields in the equations of motion. Explicitly, we have:
\begin{align}\label{LT0}
 {\cal L}_T^0&= \frac{a^3}{2}\left[T_{hh+}(\delta h)^2 + 2T_{hh\times }\delta h^i_{j}\delta h^j_{i}\right] + \bar{T}\delta_2\sqrt{|h|}+ a^3\left[ \frac{1}{2}T_{SS}\bar{h}_{ij}\delta N^i\delta N^j \right. \nonumber\\
 &\left.  +\frac{1}{2}T_{NN}(\delta N)^2 +T_{Nh}\delta N \delta h + T_{N} \left(\delta_2 N+\delta N \frac{ \delta\sqrt{|h|}}{a^3}\right)\right],
\end{align}
\begin{align}\label{LT1}
{\cal L}_T^1 &=a^3\left[ T_{\partial Sh+}\partial_i \delta N^i \delta h + 2L_{\partial Sh\times } \delta h_{ij}\partial^i \delta N^j + 2L_{hK\times} \delta h^j_{i} \delta K^i_{j} + L_{hK+} \delta K\delta h\right.\nonumber\\
& \left. + T_{NK}\delta N \delta K + T_{N\partial S}\delta N \partial_i \delta N^i \right] ,
\end{align}
\begin{align}\label{LT2}
 {\cal L}_T^2&= a^3\left[ 2L_{hR\times} \delta h^j_{i}\delta R^i_{j} + T_{hR+ } \delta R\delta h + L_R \delta_2 R +T_{\partial \dot{S}h+}\partial_i \delta \dot{N}^i\delta h  + 2L_{\partial \dot{S}h\times }\delta h_{ij}\partial^i \delta \dot{N}^j  \right. \nonumber\\
&+ \frac{1}{2}L_{\dot{S}\dot{S}}\bar{h}_{ij} \delta \dot{N}^j \delta \dot{N}^i + \frac{1}{2}T_{\partial S\partial S+} (\partial_i \delta N^i)(\partial_j \delta N^j)  + \frac{1}{2}T_{\partial S\partial S \times}\bar{h}_{lj}(\partial_i \delta N^l)(\partial^i\delta N^j) \nonumber\\
& + \frac{1}{2}L_{KK+}(\delta K)^2 + L_{KK\times}\delta K^{i}_j \delta K^j_i + \frac{1}{2}L_{\dot{N}\dot{N}}(\delta \dot{N})^2 + \frac{1}{2}T_{\partial N \partial N} \partial^i \delta N \partial_i\delta N \nonumber \\ 
&  + T_{h\partial^2N+}\delta h\partial^2 \delta N + 2L_{h\partial^2N\times}\delta h_{ij}\partial^i\partial^j \delta N + T_{NR}\delta N \delta R + L_{\dot{N}K}\delta K\delta \dot{N}  \nonumber\\
& \left.  +T_{N\partial \dot{S}} (\partial_j\delta \dot{N}^j) \delta N + L_{\partial S K+}\delta K\partial_i \delta N^i + 2L_{\partial S K\times } \delta K^i_{j}\partial_i \delta N^j \right] ,
\end{align}
where, for simplicity, we have integrated by parts, grouped coefficients together and relabeled them as $T_{*}$ (a dictionary that translates between $L_{*}$ and $T_{*}$ can be found in Appendix \ref{newcoefficients}). To understand the derivative structure above, we remind the reader that $\delta K^i_j$ contains one spatial derivative, and $\delta R$ contains two -- see the definitions in Appendix \ref{app:ADM}. For writing these actions we have also made use of the relations $\delta \sqrt{|h|}=\frac{1}{2}\sqrt{\bar h}\bar{h}^{ij}\delta h_{ij}=\frac{1}{2}a^3\delta h$ and $\delta_2 \sqrt{|h|}=\frac{1}{8}a^3(\delta h)^2-\frac{1}{4}a^3\delta h^i_{\phantom{i}j}\delta h^j_{\phantom{j}i}$. In addition, we have rewritten the term $L_K\delta K$ that comes from the expansion of $L_G$ (where now $\delta K$ includes first- and second-order perturbations). Following \cite{Gleyzes:2014rba} we have made an integration by parts so that:
\begin{eqnarray}
L_K\delta K\rightarrow -3HL_K-{\dot L}_K+{\dot L}_K\delta N+{\dot L}_K\delta_2 N -{\dot L}_K(\delta N)^2.
\end{eqnarray}
From the matter action we find the following quadratic action: 
\begin{align}\label{SecondSmv2}
S_M^{(2)}&= -\int d^4x\left\{-P_m\delta_2 \sqrt{|h|}+a^3\rho_m\left(\delta_2N+\delta N \frac{\delta_1\sqrt{|h|}}{a^3}\right)-\frac{a^3}{2}(P_m+\rho_m)(\delta N)^2 \right. \nonumber \\ 
& \left.+ a^3\left[ \frac{1}{2}V^{''}_0\delta\varphi^2 + (V'_0\delta\varphi + \dot{\varphi}_0\delta\dot{\varphi}) \delta N+ \dot{\varphi}_0\partial_i\delta \varphi\delta N^i -\frac{1}{2}\dot{\varphi}^2 + \frac{1}{2}\bar{h}^{ij}\partial_j\delta\varphi\partial_i\delta \varphi \right]\right. \nonumber \\ 
&  \left.+\delta\sqrt{|h|}\left( V'_0\delta\varphi - \delta\dot{\varphi}\dot{\varphi}_0\right)\right\}.
\end{align}
Note that $S_M^{(2)}$ leads to quadratic terms in the perturbations of the metric, as well as linear and quadratic terms on the perturbations of the matter field. We have isolated $\delta_2\sqrt{|h|}$ and $ \left(\delta_2 N+\delta N \frac{ \delta\sqrt{|h|}}{a^3}\right)$ in eq.~(\ref{LT0}), as their corresponding coefficients ($\bar{T}$ and $T_N$) will exactly cancel the corresponding terms in the matter action in eq.~(\ref{SecondSmv2}) due to the background equations (\ref{eq:back}), which can be re-expressed as:
\begin{equation}
T_{N}=\rho_m, \quad \bar{T}=-P_m,
\end{equation}
where we have used the dictionary in Appendix \ref{newcoefficients}. 

We can now apply Step 3 of the procedure described previously. First we focus only on scalar-type perturbations. We write down the total quadratic action in terms of the four metric scalars $\Phi$, $B$, $\Psi$ and $E$, and the matter perturbation field $\delta \varphi$. Note that even though we allow up to two derivatives of the metric perturbations $\delta N$, $\delta N^i$ and $\delta h_{ij}$, this means that we will have higher-order derivatives of the four scalar metric perturbations. We proceed to find the Noether identities that arise for {\it both} spatial and temporal linear gauge invariance, given in eq.~(\ref{gauge:scalarsmetric}) for the metric perturbations and analogous to eq.~(\ref{gauge:scalar}) for $\delta \varphi$. From each of these two {\it sets} of constraints (the Noether identities) we then extract the individual Noether constraints multiplying each individual perturbation and its derivatives. We solve the Noether constraints to find the following non-redundant conditions on the parameters of the quadratic action:
\begin{eqnarray}
L_{\dot{S}\dot{S}}=T_{N\partial\dot{S}}&=&
L_{K\dot{N}}=L_{\dot{N}\dot{N}}=T_{SS}=0 , \label{N0} \\
T_{NN}=T_{\partial N\partial N}&=&T_{NK}=0, \\
L_{KK+}&=&-2L_{KK\times}, \label{N2}\\
2{L}_{\partial\dot{S}h\times}-L_{K\partial S\times}&=&2{T}_{\partial\dot{S}h+}-L_{K\partial S+}=0, \\
T_{N\partial S}&=&3HL_{K\partial S+}+2HL_{K\partial S\times}, \\
2(L_{h\partial S\times}-L_{hK\times})&=&\dot{L}_{K\partial S\times}+3HL_{K\partial S\times}, \\ 
2(T_{h\partial S+}-L_{hK+})&=&\dot{L}_{K\partial S+}+ 3HL_{K\partial S+}, \\
T_{\partial S\partial S\times}+T_{\partial S\partial S +}&=&2L_{K\partial S+}+4L_{K\partial S\times} ,\\
T_{Nh}&=&3HL_{hK+}+2HL_{hK\times} , \\
2(T_{hh+}+T_{hh\times})&=&\dot{L}_{hK+}+\dot{L}_{hK\times}+3HL_{hK+}+3HL_{hK\times}, \\
4T_{hR+}&=&L_R+L_{KK\times}+\dot{L}_{KK\times}/H, \\
2L_{hh\times}&=&\dot{L}_{hK\times}+3HL_{hK\times} , \label{N1}\\
H(L_R-4L_{hR\times}) &=&\dot{L}_{KK\times}+HL_{KK\times}, \\
T_{h\partial^2 N+}&=&-2L_{h\partial^2 N\times}, \\
T_{NR} &=& L_{KK\times}-2L_{h\partial^2N\times}, \\
4\dot{H}L_{KK\times}&=&-(\rho_m+P_m), \label{N3} 
\end{eqnarray}
where we have used the background equations to simplify some of these constraints. We have written these equations in a form that will look the same for a minimally coupled scalar field and a general perfect fluid. Notice that all these constraints can be solved algebraically, by simply working out one coefficient without time derivatives in terms of the rest.

Via the constraints above, the number of free coefficients in our original action is greatly reduced. A straight substitution of the Noether constraints into the quadratic action reduces the original 32 free, time-dependent,
functions (30 coefficient functions $L_*$ and $T_*$ in $S_G^{(2)}$, along with the two background functions $\varphi_0$ and $a$) down to 8; after some integrations by parts we can collapse the number of the remaining free functions down further to only one: $L_{KK\times}$. In addition, we find that all terms involving time derivatives of $\delta N$ and $\delta N^i$ vanish, so they play the role of functional Lagrange multipliers -- one of the key characteristics of General Relativity. 

It is apparent that the time dependence of the coefficients is intimately tied to that of the background, through $H$, $\rho_m$ and $P_m$. We can then take the final step of replacing our reduced set of coefficients into eq.~(\ref{GRaction}), to get the following total quadratic action: 
\begin{eqnarray}
S_G^{(2)}+S_M^{(2)}&=&\int d^4 x\; a^3\left[-\dot{\varphi}_0
\delta\dot{\varphi}\left(\Phi+3\Psi\right) -V'
\delta\varphi\left(\Phi-3\Psi\right) -\frac{1}{2}V''\left(\delta\varphi\right)^2-\frac{1}{2}\delta\varphi\partial^2\delta\varphi\right. \nonumber \\&+&\left. \frac{1}{2}\left(\delta\dot{\varphi}\right)^2
-\dot{\varphi}_0a^2\partial^2E\delta\dot{\varphi}-\dot{\varphi}_0\delta\varphi\partial^2B+V'\delta\varphi a^2\partial^2E
+M^2\left(1+\frac{d\ln M^2}{d\ln a}\right)
\Psi\partial^2\Psi
\right. \nonumber \\
&-&\left. 3M^2\dot{\Psi}^2-6HM^2\dot{\Psi}\Phi-2M^2\Psi\partial^2\Phi 
 -\left(\dot{H}+3H^2\right)M^2\Phi^2
 \right. \nonumber \\
&-&\left. 2M^2a^2\partial^2\dot{E}\left(\dot{\Psi}+H\Phi\right)
 +2M^2\dot{\Psi}\partial^2B+2HM^2\Phi\partial^2B
 \right],
\end{eqnarray}
where we have redefined $M^2\equiv2L_{KK\times}$, so that one of the Noether constraints becomes,
\begin{eqnarray}
M^2=-\frac{\rho_m+P_m}{2\dot{H}}. \label{M2eq}
\end{eqnarray} 

It is instructive to further transform this quadratic action, by making the replacement $\delta\varphi \rightarrow \delta\varphi \dot{\varphi}_0$, using the background equation for the scalar field (eq.~(\ref{N4})), and making a few integrations by parts. We then find:
\begin{eqnarray}
S_G^{(2)}+S_M^{(2)}&=&\int d^4 x\;  a^3M^2\left[-6\dot{H}\delta\varphi\dot{\Psi}
-\dot{H}\left(\delta\dot{\varphi}\right)^2
+2\dot{H}\Phi\delta\dot{\varphi}
-2\dot{H}a^2\partial^2\dot{E}\delta\varphi
\right.
\nonumber \\
 &-& \left. 3\dot{H}^2\left(\delta\varphi\right)^2
-6H\dot{H}\delta\varphi\Phi
+2\dot{H}\delta\varphi\partial^2B
+\dot{H}\delta\varphi\partial^2\delta\varphi
+ \left(1+\frac{d\ln M^2}{d\ln a}\right)\Psi{ \partial}^2\Psi
\right.
\nonumber \\ 
&-& \left. 3{\dot \Psi}^2 
-6H\Phi{\dot \Psi}
-2\Psi{\partial}^2\Phi-\left(\dot{H}+3H^2\right)\Phi^2\right.
\nonumber \\ 
&-& \left. 2a^2\partial^2{\dot E}\left({\dot \Psi}+H\Phi\right)
+2{\dot \Psi}{ \partial}^2B 
+2H\Phi{\partial}^2B \right].
\label{GR}
\end{eqnarray}

In other words, the final action in terms of the metric perturbations depends only on one free function of time, $M^2$; the scale factor does not count as a free function, as it is related to $M$ through eq.~(\ref{M2eq}) and (\ref{N4}). If the background equations were simply the Friedman equations then, from eq.~(\ref{M2eq}) we would find $M^2=M^2_{\rm Pl}$, and eq.~(\ref{GR}) would become the quadratic action for General Relativity. In general, however, $M^2$ is a completely free function of time. 
This illustrates a crucial feature of any approach based on finding general linearized theories at the perturbative level. For a single tensor, at the level of the \textit{full} diffeomorphism-invariant theory, we know that there should be no overall free function of time left -- GR is unique in this sense. Said another way, $M^2$ being a free function of time is an artefact of just taking into consideration the linearized action for perturbations. The consistency of a full theory requires background, linearized perturbative and higher-order perturbative contributions all to be consistent, i.e.~to avoid the propagation of unstable degrees of freedom such as ghosts. And so, crucially, while all well-behaved theories will map onto the free functions in our linearized perturbation theory parametrization, not all possible functional forms for these seemingly free functions are associated with healthy theories. This happens for the very simple reason that there is more to a full theory than the action it gives rise to for linear perturbations, and that there are additional constraints not captured by any formalism based on linearized perturbations. These extra constraints will reduce the free functions we recover further. A detailed analysis on the construction of possible fundamental consistent theories leading to the quadratic actions presented here is beyond the scope of this paper, but it is certainly relevant and requires further work.

We have shown that it is possible to systematically recover the linearized action for the most general linearly diffeomorphism-invariant theory of gravity built from a metric, by starting from a completely general action and systematically applying gauge transformations to obtain the Noether constraints. 
We have found that $M^2$ is the only parameter that enters the final action and hence the equations of motion. Looking forward, this means that any attempt to constrain this action 
(with cosmological observations) boils down to constraining $M^2$. But, as we have seen, there are a number of degeneracies that remain between the original coefficients $L_*$. So, we can already see that it is impossible to individually constrain all the coefficients that we used to build the
action in equations (\ref{LT0})-(\ref{LT2}). In effect, we will never be able to completely pin down the landscape of theories to solely GR using only cosmological linear perturbation theory alone. At best we will be able to constrain these actions to a degenerate family of theories that includes GR.

Finally, we remark that since Action (\ref{GR}) leads to, at most, second-order differential equations in time, it is free of Ostrogradski instabilities associated to higher time-derivative terms \cite{Clifton:2011jh,Woodard:2006nt}. Furthermore, this action propagates only one physical scalar DoF, which actually comes from the matter sector. It can be seen that $\Phi$ and $B$ are auxiliary variables, i.e.~without dynamics, and can be expressed in terms of the rest of the fields by using their own equations of motion. Therefore, they do not represent independent physical DoFs. In addition, the action has a gauge symmetry with two arbitrary parameters inducing two redundant fields in the action. Thus, from the original 5 scalars in eq.~(\ref{GR}), only one field is physical. 

\section{Recovering linearized Beyond Horndeski theory and beyond}
\label{sec:ST}

The simplest, non-trivial example of a theory which includes an extra degree of freedom and differs from General Relativity is a scalar-tensor theory. The original, most elementary, formulation is Jordan-Brans-Dicke gravity, a theory in which the Planck mass is promoted to a dynamical scalar field \cite{Jordan:1959eg,Brans:1961sx,Brans:1962zz}. Jordan-Brans-Dicke gravity has been one of the workhorses of modern cosmology and has been deployed in understanding both the early universe (specifically inflation) and the late-time accelerated expansion of the Universe \cite{Clifton:2011jh}. Over the past few years, renewed interest in scalar-tensor theories has emerged, on the one hand from the rediscovery of the Horndeski action \cite{Horndeski:1974wa} -- the most general, non-degenerate, scalar-tensor action with second order equation of motion -- and on the other hand from various extensions of the class of covariant Galileons \cite{Deffayet:2009wt}. 

As mentioned in the introduction, most attempts at constructing a general parametrization of linearized gravity have focused on scalar-tensor theories. A nuanced understanding of how scalar-tensor theories emerge has been developed, most notably in \cite{Bellini:2014fua}, where an economical parametrization of such theories was proposed in terms of four free functions. These functions (the `$\alpha$' functions) can be easily related to specific physical properties of the fundamental action. Subsequent work has extended this parametrization to five free functions \cite{Gleyzes:2014dya,Gleyzes:2014qga,Zumalacarregui:2013pma}. 

In what follows, we will parametrize linearly diffeomorphism-invariant gravitational theories containing one metric and one scalar field, coupled minimally to a matter scalar field (although the results presented here also hold for a general matter perfect fluid). As in the previous section, we will analyze linear perturbations of the fields around a homogeneous and isotropic background. We will show that with our procedure we can reproduce previous work. In particular, we will show how the free functions describing such theories will emerge from the Noether constraints applied to a quadratic action with up to three time and space derivatives. Furthermore, we will then show that, if we include higher-order derivatives, a further set of functions must be included to completely cover the possible space of theories. To avoid any Ostrogradski instability, we allow at most two \textit{time} derivatives of the fields, but higher \textit{spatial} derivatives are permitted (this situation can arise in some Lorentz-violating theories, but also in some special Lorentz-invariant cases such as Beyond Horndeski theories). It would of course be possible to go beyond this and find theories that are higher-order in temporal derivatives as well, yet evade Ostrogradski ghosts via the presence of degeneracies \cite{Langlois:2015cwa,Achour:2016rkg} or, equivalently, hidden constraints \cite{Crisostomi:2016czh}. \footnote{In this context also note that we are interested in effective theories, which should be ghost-free within their regimes of validity. Any given model may ``predict'' ghost-like instabilities outside the regime of validity of that theory, i.e.~instabilities coming with a mass/energy scale above the theory's cutoff. However, such instabilities are not physical and there is no reason to discard a theory.}  

\subsection{Horndeski and beyond}
We now include an extra degree of freedom, a scalar field $\chi$, whose perturbation transforms under linear coordinate transformations as in eq.~(\ref{gauge:scalar}). We proceed with Step 2 for constructing the most general quadratic action. Allowing at most three derivatives of the perturbation fields (two temporal but three spatial), we write down all possible perturbed building blocks $\delta\vec{\Theta}= (\cdots, \delta \chi, \delta \dot{\chi}, \partial_i\delta \chi, \partial_i\delta \dot{\chi}, \partial_i\partial_j\delta \chi,\partial_i\partial_j\delta\dot{ \chi}, \partial_i\partial_j\delta_k\delta \chi , \delta\partial_i\partial_j \dot{N},\delta\partial_i\partial_j \partial_k  N,$\\ 
$\delta \partial_i\partial_j \dot{N}^k,\delta \partial_i\partial_j \partial_k N^l)$ where the initial ellipses indicates all the building blocks used in Section \ref{sec:GR} \footnote{As in the previous section, we do not consider a term with three spatial derivatives of $\delta h_{ij}$. Such terms could be added but the form of the quadratic Lagrangians ${\cal L}_T^i$ and ${\cal L}_\chi^i$ would not change except in terms of the explicit relations between the coefficients $T_*$ and $L_*$.}. We will also introduce more definitions for the coefficients $L_*$, in addition to those given in eq.~(\ref{DefL2D}) and (\ref{DefL2D2}):
\begin{align}
 L_{A_i B_{jkl}}&=\frac{1}{3}L_{AB}\left(\bar{h}^{ij}\bar{h}^{kl}+\bar{h}^{ik}\bar{h}^{jl}+\bar{h}^{il}\bar{h}^{jl}\right), \; \mbox{where $B_{ijk}$ is fully symmetric},  \nonumber \\
 L_{A^i_{\phantom{i}jkl}}&=\frac{1}{3}L_A \left(\delta^{j}_{\phantom{j}i}\bar{h}^{kl}+\delta^{k}_{\phantom{k}i}\bar{h}^{jl}+\delta^{l}_{\phantom{l}i}\bar{h}^{jl}\right), \; \mbox{where $A^i_{\phantom{i}jkl}$ is symmetric in 3 indices}, \nonumber\\
 L_{BA^i_{\phantom{i}jkl}}&=\frac{1}{3}L_{BA}\left(\delta^{j}_{\phantom{j}i}\bar{h}^{kl}+\delta^{k}_{\phantom{k}i}\bar{h}^{jl}+\delta^{l}_{\phantom{l}i}\bar{h}^{jl}\right), \; \mbox{where $A^i_{\phantom{i}jkl}$ is symmetric in 3 indices},
\end{align}
where $A_i$, $B_{jkl}$, etc.~correspond to any possible building block with the corresponding index structure. An exceptional case is 
\begin{align}
 L_{h_{ij}\partial_k\partial_l\partial_m N^n} \delta h_{ij}\partial_k\partial_l\partial_m \delta N^n &= L_{h\partial^3 S+}\delta h\partial^2\partial_i \delta N^i + 2 L_{h\partial^3 S\times}\delta h_{ij}\partial^2 \partial^i\delta N^j \nonumber\\
& + 2L_{h\partial^3 S\odot}\delta h_{ij}\partial^i\partial^j \partial_l\delta N^l.
\end{align}

As in the previous section, we Taylor expand the gravitational and matter action up to second order in the perturbation fields. From the linear total action we derive the background equations. If we do so, we will obtain eqs.~(\ref{eq:back}) and eq.~(\ref{N4}) for the metric evolution and matter field, which we now supplement with: 
\begin{equation}
L_{\chi} - 3H L_{\dot{\chi}} - \dot{L}_{\dot{\chi}}=0 \label{eq:chi0},
\end{equation}
which corresponds to the background equation for the scalar field $\chi_0$. These four background equations should not be all independent, as there are only three undetermined background functions: $a$, $\chi_0$ and $\varphi_0$. This redundancy imposes a relation between the coefficients $T_*$ and $L_*$, which is not relevant for this paper, but would be important for the task of constructing non-perturbative fundamental actions allowing homogeneous and isotropic backgrounds. 

We now extend the gravitational action considered in Section \ref{sec:GR} such that
\begin{equation}
S_G^{(2)}=\int d^4 x\;  \sum_{i=0}^3 \left({\cal L}_T^{i}+ {\cal L}_\chi^{i}\right) , 
\end{equation}
where ${\cal L}_\chi^{i}$ are quadratic Lagrangians involving $\delta \chi$, leading to $i$ derivatives of the perturbation fields in the equations of motion. Up to second-order derivatives, we have the tensor Lagrangians given in the previous section, and we add the following Lagrangians involving the perturbation of the scalar field, $\delta\chi$:
\begin{eqnarray}
{\cal L}_\chi^{0}&=& \frac{a^3}{2}\left[T_{\chi\chi}(\delta \chi)^2+2T_{\chi h}\delta \chi \delta h+ T_{\chi N } \delta N \delta \chi\right], \\
{\cal L}_\chi^{1}&=& a^3\left[ T_{\dot{\chi}h }\delta \dot{\chi }\delta h + L_{\chi K}\delta \chi \delta K +T_{\chi\partial S} \delta \chi \partial_i\delta N^i + T_{\dot{\chi}N } \delta \dot{\chi }\delta N \right], \\
{\cal L}_\chi^{2}&=& a^3\left[ L_{\chi R}\delta \chi \delta R +T_{\partial^2 \chi h+ }\delta h\partial^2 \delta \chi + 2L_{\partial^2 \chi h\times}\delta h_{ij}\partial^i\partial^j\delta \chi + \frac{1}{2}L_{\dot{\chi}\dot{\chi}} (\delta \dot{\chi})^2 + L_{K\dot{\chi}}\delta K \delta \dot{\chi} \right.\nonumber\\ 
&+& \left. \frac{1}{2}T_{\partial\chi\partial\chi }\partial_i\delta\chi \partial^i\delta \chi + L_{\dot{\chi}\dot{N}}\delta\dot{N} \delta \dot{\chi}+ T_{\dot{\chi}\partial S}\partial_i\delta N^i\delta \dot{\chi} + T_{\partial\chi\partial N }\partial_i\delta N\partial^i\delta \chi\right],\label{Lchi2}
\end{eqnarray}
For third-order derivatives we include the following tensor and scalar Lagrangians:
\begin{eqnarray}
 {\cal L}_T^3 &=& a^3\left[ 2L_{ h\partial^3 S\times }\delta h_{ij}\partial^2\partial^i\delta N^j + T_{h\partial^3 S+ } \delta h\partial^2 \partial_j \delta N^j \right. \nonumber \\
 &+&  2L_{h\partial^3 S\odot}\delta h_{ij}\partial^i\partial^j\partial_l\delta N^l + L_{h\partial^2 \dot{N}+} \delta h \partial^2 \delta \dot{N} + 2L_{h\partial^2 \dot{N}\times }\delta h_{ij}\partial^i \partial^j \delta \dot{N} + L_{\partial SR+}\delta R \partial_j \delta N^j \nonumber\\
&+&  2L_{\partial S R\times} \delta R^i_{j}\partial_i\delta N^j + L_{KR+} \delta K\delta R + 2L_{KR\times}\delta K^i_j\delta R^j_i +L_{\dot{N}R}\delta R \delta \dot{N} + T_{K\partial^2 N + } \delta K \partial^2 \delta N \nonumber\\
&+&  2L_{K\partial^2 N \times}\delta K^i_j \partial^j\partial_i \delta N +L_{\partial\dot{S}K+} \delta K\partial_i \delta \dot{N}^i + 2L_{\partial\dot{S}K\times} \delta K^i_{j}\partial_i \delta \dot{N}^j + T_{\dot{N}\partial \dot{S} } \delta \dot{N}\partial_j \delta \dot{N}^j \nonumber\\
&+&\left. T_{\partial^2 N\partial S } \partial_i \delta N^i \partial^2 \delta N \right],
\end{eqnarray}
and
\begin{eqnarray}
{\cal L}_\chi^{3}&=&a^3\left[L_{R\dot{\chi}}\delta R \delta \dot{\chi} +T_{\partial^2\dot{\chi}h+ }\delta h\partial^2 \delta \dot{\chi} + 2L_{\partial^2\dot{\chi}h\times}\delta h_{ij}\partial^i\partial^j\delta \dot{\chi} + T_{\partial^2\dot{\chi}N } \delta N\partial^2\delta \dot{\chi} + T_{\dot{\chi}\partial \dot{S}}\partial_i \delta \dot{N}^i\delta \dot{\chi} \right. \nonumber\\
 &+& \left.  L_{\partial^2 \chi K+} \delta K\partial^2\delta \chi + 2L_{\partial^2\chi K\times}\delta K^i_j\partial^j\partial_i\delta \chi + T_{\partial^2 \chi \partial S }\partial_i \delta N^i \partial^2 \delta \chi \right].
\end{eqnarray}
Note that, as in Section \ref{sec:GR}, we have integrated by parts and redefined some of the coefficients to simplify notation; the dictionary to translate between the $T_*$ and $L_*$ is in Appendix \ref{app:newcoeffscalar}. Finally, for the matter scalar field $\varphi$ we add the quadratic action shown in eq.~(\ref{SecondSmv2}).

We now follow Step 3 where, in addition to  (\ref{N0}), (\ref{N2})-(\ref{N1}) we get a new set of Noether constraints. 
We find that the end result depends on five free coefficients of the time: $M^2$, $\alpha_B$, $\alpha_K$, $\alpha_T$, $\alpha_H$. In terms of the coefficients $L_*$ and $T_*$, arising solely in ${\cal L}^i_T$, these are:
\begin{eqnarray}
M^2&= &2L_{KK\times}, \label{DefM2v2}\\
\alpha_B&=&\frac{1}{2}\frac{T_{NK}}{M^2H}, \label{DefAlphaB}\\
\alpha_K&=&\frac{T_{NN}+T_{SS}}{H^2M^2},\\
\alpha_T&=&\frac{2}{M^2}\left(L_R+HL_{KR\times}+\dot{L}_{KR\times}+4L_{Rh+} \right)-1,\\
\alpha_H&=&
\frac{2}{M^2}\left[-\dot{L}_{K\partial^2N\times}+H\left(L_{KR\times}-L_{K\partial^2N\times}\right)+T_{NR}+2L_{h\partial^2N\times}\right]-1 ,\label{alphahorndeski}
\end{eqnarray}
which are completely equivalent (and more general) than the expressions found in \cite{Gleyzes:2014rba}. Note that the $\alpha_i$ can be neatly understood via the physical effects they parametrize \cite{Bellini:2014fua}. Explicitly, the final quadratic gravitational action is then:
\begin{eqnarray}\label{SFourthOrderFinal}
S_G^{(2)}&=&\int d^4x\; a^3M^2\left\{\frac{1}{2}H^2\left(\alpha_K-12\alpha_B-6\right)\Phi^2-6H\left(1+\alpha_B\right)\Phi\dot\Psi
+2\left(1+\alpha_H\right)\Psi\partial^2\Phi
\right. \nonumber \\ 
&-& \left. 3\dot{\Psi}^2-\left(1+\alpha_T\right)\Psi\partial^2\Psi
+2a^2H\left(1+\alpha_B\right)\Phi\partial^2\dot{E}-2H\left(1+\alpha_B\right)\Phi\partial^2B + 2a^2\Psi\partial^2\dot{E}
\right. \nonumber \\ 
&-& \left. 2\dot{\Psi}\partial^2B
-3\left(\frac{\rho_m+P_m}{M^2}+2\dot{H}\right)\dot{\Psi}\delta \chi + 2\alpha_H\dot{\Psi}\partial^2\delta\chi +6H\alpha_B\dot{\Psi}\delta\dot{\chi} + H^2\left(6\alpha_B-\alpha_K\right)\Phi\delta\dot{\chi}
\right. \nonumber \\ 
&-& \left. 2H\left[\alpha_T-\alpha_H-\frac{d\ln M^2}{d\ln a}\left(\alpha_H+1\right)-\frac{d\alpha_H}{d\ln a}\right]\Psi\partial^2\delta\chi -2H(\alpha_B-\alpha_H)\Phi\partial^2\delta\chi
\right. \nonumber \\ 
&-& \left. 3H\left[\frac{\left(\rho_m+P_m\right)}{M^2}+2\dot{H}\left(1+\alpha_B\right)\right]\Phi\delta\chi - \left[\frac{\left(\rho_m+P_m\right)}{M^2}+2\dot{H}\right]\delta\chi\left(\partial^2B-a^2\partial^2\dot{E}\right)
\right. \nonumber \\ 
&+& \left. 2H\alpha_B\delta\dot{\chi}\left(\partial^2B-a^2\partial^2\dot{E}\right) - \left[3\left(\dot{H}^2+H\ddot{H}+3H^2\dot{H}+H^2\dot{H}\frac{d\ln M^2}{d\ln a}\right)\alpha_B + 3H\dot{H}\dot{\alpha}_B\right. 
\right. \nonumber \\ 
&+& \left. \left. \frac{3}{2}\dot{H}\frac{\left(\rho_m+P_m\right)}{M^2}+3\dot{H}^2\right]\delta\chi^2 - \left[\left(\dot{H}+H^2+H^2\frac{d\ln M^2}{d\ln a}\right)\left(\alpha_B-\alpha_H\right) + H\left(\dot{\alpha}_B- \dot{\alpha}_H\right)\right. \right. \nonumber \\ 
&+& \left. \left. H^2\alpha_T+\dot{H}- H^2\frac{d\ln M^2}{d\ln a}+\frac{1}{2}\frac{(\rho_m+P_m)}{M^2}\right]\delta\chi\partial^2\delta\chi
+\frac{1}{2}H^2\alpha_K\delta\dot{\chi}^2\right. \\
&-&\left. P_m\left( \frac{3}{2}\Psi^2-a^2\Psi\partial^2 E -\frac{a^4}{2}\partial^2 E\partial^2E \right)-\rho_m\left(\frac{1}{2}\Phi^2+\frac{1}{2} B\partial^2 B +3\Phi \Psi- a^2 \Phi \partial^2 E\right) \right\},\nonumber
\end{eqnarray}
where we have redefined $\delta \chi \rightarrow \delta\chi \dot{\chi}_0$. Note that all the terms in the last line are those arising from $\delta_2\sqrt{|h|}$ and $(\delta_2 N+\delta\sqrt{|h|}\delta N)$, and they will all cancel with an equivalent counterpart from the matter action $S_M^{(2)}$. Given that the background depends on two free functions $a$ and $\chi_0$ ($\varphi_0$ is not free as it will be related to $a$ by means of eq.~(\ref{N4})), we have shown that this cosmological model is completely characterized by seven free functions of time, parametrizing the evolution of the background and linear perturbations. Note that, in this case we do not have any extra relation such as eq.~(\ref{M2eq}) relating the background functions to $M$. We emphasize that, even though we did our calculations with a matter scalar field, our expression for $S_G^{(2)}$ is valid when the matter sector is a general perfect fluid instead of a scalar field. The equations of motion for this gravitational model coupled to a general perfect fluid can be derived from equations (\ref{fulleom}).

The action we have just determined includes up to third-order derivatives of the perturbation fields. The coefficients $M^2$, $\alpha_K$, $\alpha_B$ and $\alpha_T$ multiply terms that have, at most, two derivatives, and therefore encompass fundamental theories such as Horndeski theory. But we also found a ``beyond Horndeski" coefficient, $\alpha_H$, which multiplies a term of the form $\dot{\Psi}\partial^2\delta\chi$ that has three derivatives; therefore our results encompass the extensions from Beyond Horndeski theory. 

We can recover the results of the previous section by setting $\chi=0$. The free coefficients then take the following values:
\begin{equation}
\alpha_{K}=\alpha_{B}=\alpha_H=0, \quad \alpha_T=\frac{d\ln M^2}{d \ln a}, 
\end{equation}
which corresponds to GR if $M$ is constant ($\alpha_T=0$). On the other hand, if we want $\delta \chi$ to describe the perturbations of a quintessence scalar, we set the coefficients to be:
\begin{equation}
\alpha_{B}=\alpha_H=0, \quad \alpha_T=\frac{d\ln M^2}{d \ln a}, \quad \alpha_{K}=\frac{\dot{\chi}_0^2}{H^2M^2}.
\end{equation}
Note that by constraining the form of the terms for $\delta \chi$ in this way, we are also constraining the quadratic tensor terms -- they are all related. In this case, the tensor action reduces to that of the generalized GR action shown in the previous section. If we restrict ourselves to GR, we find $\alpha_T=0$ as in \cite{Bellini:2014fua}\footnote{Note that there is a typo in Table 1 of \cite{Bellini:2014fua} - a factor of $3$ is missing in the definition of $\alpha_K$ for quintessence.}.

We finally comment on the fact that action (\ref{SFourthOrderFinal}) propagates only one physical scalar DoF. It can be seen that $B$ and $\Phi$ are auxiliary variables, while the other three fields have time derivatives; and due to the redundancies induced by the two scalar gauge freedoms, the action contains only one physical, propagating, scalar DoF. 

\subsection{Fourth-order extensions}
It is interesting to go further to see what the structure of higher-order derivative terms might take and what new free coefficients must be included. Both \cite{Gubitosi:2012hu} and \cite{Bloomfield:2012ff} include a term of the form $\left(g^{\mu\nu}+n^\mu n^\nu\right)\partial_\mu g^{00}\partial_\nu g^{00}$ in the unitary gauge which, when Stueckelberg-ed, leads to a {\it fourth-order} derivative term of the scalar field in the quadratic action of the form $\alpha_P\partial^i\delta\dot{\chi}\partial_i\delta\dot{\chi}$ where $\alpha_P$ can be expressed as
\begin{eqnarray}
\alpha_P= \frac{\dot{\chi}^2_0T_{\partial\dot{\chi}\partial\dot{\chi}}}{M^2H^4a^2},
\end{eqnarray}
and $T_{\partial\dot{\chi}\partial\dot{\chi}}$ is the coefficient in the quadratic action multiplying a term of the form $\partial^i\delta\dot{\chi}\partial_i\delta\dot{\chi}$.
More recently, in \cite{Langlois:2015cwa,Achour:2016rkg, Crisostomi:2016czh}, the authors explored the possibility of enlarging the family of viable scalar-tensor theories by allowing fourth-order derivatives of the scalar field in the equations of motion, but avoiding Ostrogradski instabilities through additional (hidden) constraints. 

We now go beyond ``Beyond Horndeski", to see what kinds of terms arise by systematically including {\it all} possible fourth-order derivative terms in the quadratic action (i.e.~including Lagrangians ${\cal L}_T^4+{\cal L}_\chi^4$, with up to four spatial derivatives but only two time derivatives). We find that the final action now depends on the five coefficients previously found as well as six {\it new} coefficients, one of which is the $\alpha_P$ found in \cite{Bloomfield:2012ff}.
The new coefficients are defined in the following way:
\begin{align}
\alpha_{Q1}&=\frac{H^2}{2M^2}\left(4L_{RR+}+ 3L_{RR\times}\right),\\
\alpha_{Q2}&=\frac{2}{M^2}\left(L_{KK+}+2L_{KK\times}\right), \\
\alpha_{Q3}&=\frac{H}{M^2}\left(L_{KR+}+L_{KR\times}\right),\\
\alpha_{Q4}&=\frac{H}{M^2}\left(T_{\partial^2N\partial^2\chi}-\frac{2}{3}L_{K\partial^2\dot{\chi}\times} \right)\dot{\chi}_0,\\
\alpha_{Q5}&=\frac{H}{M^2} T_{\partial^2N\partial^2\chi} \dot{\chi}_0, \\
\alpha_{P}&= \frac{\dot{\chi}^2_0}{M^2H^4a^2} T_{\partial\dot{\chi}\partial\dot{\chi}}.
\end{align}
(Note that, as for equations (\ref{DefM2v2})-(\ref{alphahorndeski}), we could rewrite all these new coefficients in terms of $L_*$ and $T_*$ solely from the tensor part of the action but the expressions would be more cumbersome). These terms contribute with the following fourth-order derivative interaction terms to the final quadratic action (as well as contributing to lower-order derivative terms): 
\begin{align}
&\alpha_{Q1}\rightarrow \{\partial^2\Psi\partial^2\delta \chi, \partial^2 \delta\chi \partial^2 \delta\chi, \partial^2 \Psi \partial^2 \Psi  \},\\
&\alpha_{Q2}\rightarrow \{\partial^2  \delta\chi \partial^2  \delta\chi \},\\
&\alpha_{Q3}\rightarrow \{	\partial^2\delta\chi \partial^2\delta\chi , \partial^2 \delta\chi \partial^2\Psi \},\\
&\alpha_{Q4}\rightarrow \{\partial^i \delta\dot{\chi}\partial_i\dot{\Psi}\}, \\
&\alpha_{Q5}\rightarrow \{\partial^2\delta\dot{\chi}\partial^2 \dot{E}, \partial^2\delta\chi \partial^2\delta\chi ,\partial^2\delta\chi\partial^2 \Phi, \partial^2\delta\dot{\chi}\partial^2 B \}, \\
&\alpha_{P}\rightarrow \{\partial^i\delta\dot{\chi} \partial_i\delta\dot{\chi}\}.
\end{align}
Notice that all these terms have four derivatives of the perturbation fields $\delta h_{ij}$, $\delta N$, $\delta N^i$ and $\delta \chi$, but when using the SVT decomposition they have higher derivatives of the scalar-type perturbations. For completeness we list ${\cal L}_T^4$ and ${\cal L}_\chi^4$ in Appendix \ref{4thorderaction}.

 The final quadratic action is lengthy, but can be found explicitly in the {\it xIST} notebook {\it COPPER}, published together with this paper (see link in Section~\ref{sec:intro}). Although this final action becomes more complex, it has the same structure that we see in the action of equation (\ref{SFourthOrderFinal}): $B$ and $\Phi$ are auxiliary variables, while the other three fields have time derivatives; and after using the gauge freedom, the action contains only one physical, propagating, scalar DoF. The quadratic actions found here with four derivatives should encompass some specific cases of the scalar-tensor theories considered in \cite{Langlois:2015cwa,Achour:2016rkg,Crisostomi:2016czh}.

It is important to remark that some scalar-tensor actions could have a different structure and allow the quadratic term $\dot{\Phi}^2$. As shown in \cite{Gleyzes:2014rba}, such actions could be obtained by performing a conformal transformation of the metric with a dependence on derivative terms of the scalar field $\chi$ to the action in eq.~(\ref{SFourthOrderFinal}). Even though $\Phi$ would not be an auxiliary field anymore, these actions would propagate the same number of DoFs as the actions found in this section, due to the presence of additional (hidden) constraints. We do not find the term $\dot{\Phi}^2$ in our results because the presence of such term requires the presence of other quadratic terms (in order to have a gauge-invariant action) of the form $\ddot{\chi}^2$ that lead to four time derivatives in the equations of motion, which we ignored. Furthermore, in \cite{Crisostomi:2016czh} it was shown explicitly that after conformal transformations with kinetic dependence on the scalar field, the action of Horndeski is mapped into a specific action that leads to fourth derivatives in the equations of motion. Thus, we emphasize that the absence of these terms in our results does not represent a restriction on the formalism but on the extra assumptions made for the specific cases we worked out instead. In fact, if we had allowed four time derivatives of the fields, we would have found the term $\dot{\Phi}^2$ in the final quadratic action.


\section{Vector-Tensor theories}
\label{sec:vector}

In previous sections we have focused on scalar-tensor modified gravity theories; in this section we show how the method can easily be extended to vector-tensor gravity theories. Vector-tensor theories have been studied in detail in attempts to understand spontaneous Lorentz violation \cite{Jacobson:2000xp,Zlosnik:2007bu}, to generate massive gravitons \cite{Gripaios:2004ms} and as models of dark matter and dark energy \cite{Zuntz:2010jp,Zlosnik:2006zu}.

\subsection{General case}

We aim to parametrize linearly diffeomorphism-invariant quadratic actions containing one metric and one vector field $A^\mu$. As in the previous sections, we add a scalar field, minimally coupled to the metric, to represent the matter sector, and consider linear perturbations of all the fields around a homogeneous and isotropic background. For the vector field we will have:
\begin{equation}
A^\mu=\left(A,\vec{0}\right)+ \alpha^\mu,
\end{equation}
where $A(t)$ is the background solution of the vector field, and $\alpha^\mu$ its first-order perturbation. Since we will be focusing on scalar-type perturbations, we use the SVT decomposition of the vector field to write:
\begin{equation}
\alpha^\mu=(\alpha^0, \alpha^i); \quad \alpha^i=\alpha^{Ti}+\bar{h}^{ij}\partial_j\alpha,
\end{equation}
where we have two scalar perturbations $\alpha^0$ and $\alpha$, and one vector perturbation $\alpha^{Ti}$, such that $\partial_i \alpha^{Ti}=0$. Therefore there will only be two relevant perturbations (the two scalar modes) from the vector field in our calculations. As explained in Appendix \ref{app:gauge}, these scalar perturbations transform in the following way under linear coordinate transformations:
\begin{align}
\delta \alpha^0 &= \dot{\pi}A -\dot{A}\pi, \nonumber\\
\delta \alpha &= a^2A\dot{\epsilon}, 
\end{align}
while the scalar metric perturbations transform as in eq.~(\ref{gauge:scalarsmetric}) and the matter scalar field $\delta \varphi$ as in eq.~(\ref{gauge:scalar}). 

We now follow Step 2 to construct the most general gravitational quadratic action. We will allow, at most, two derivatives of the perturbation fields in the equations of motion. All the possible perturbed building blocks in this case will be $\delta\vec{\Theta}= (\ldots, \alpha^0, \partial_i\alpha^0,$ $\dot{\alpha}^0, \partial_i\partial_j\alpha^0, \partial_i\dot{\alpha}^0,\alpha_i, \partial_j\alpha_i, \dot{\alpha}_i, \partial_j\partial_k\alpha_i, \partial_j\dot{\alpha}_i),$ where the initial ellipses indicate all the building blocks used in Section \ref{sec:GR}. For simplicity we have defined $\alpha_i=\bar{h}_{ij}\alpha^i$ which, in terms of scalar-type perturbations, becomes $\alpha_i=\partial_i\alpha$. 

Next we proceed to Taylor expand the gravitational Lagrangian $L_G$ up to second order in the perturbation fields. We use the same definitions introduced in eq.~(\ref{DefL2D}) for the coefficients $L_*$. In addition, we use $\alpha$ as a proxy for $\alpha_i$ in the subscripts of the coefficients $L_*$. We also Taylor expand the matter action. 

We recall that we obtain the background equations of motion from the linear Taylor expansion of the total action (gravity and matter). In this case, we find eq.~(\ref{eq:back}) from varying the metric field, eq.~(\ref{N4}) from the matter scalar field, and the following expression from varying the vector field:
\begin{equation}
L_{\alpha^0}-\dot{L}_{\dot{\alpha}^0}-3HL_{\dot{\alpha}^0}=0.
\end{equation}
Similar to the case of scalar-tensor theories, we expect one of these four background equations to be redundant as there are only three undetermined background functions $a$, $A$ and $\varphi_0$. Again, this redundancy leads to a relation between the parameters $L_*$ and $T_*$, which is not relevant for the analysis of this paper, but would be important in constructing non-perturbative, fundamental actions allowing homogeneous and isotropic backgrounds.  

We now proceed to express the general quadratic gravitational action as:
\begin{equation}
S_G^{(2)}=\int d^4x\; \sum_{i=0}^2  \left( {\cal L}_T^i+{\cal L}_{\alpha^0}^i+{\cal L}_{\alpha}^{i} + {\cal L}_{\alpha^0\alpha}^i\right) ,
\end{equation}
where ${\cal L}_{\alpha^0}^i$ and ${\cal L}_{\alpha}^i$ are the quadratic Lagrangians involving $\alpha^0$ and $\alpha_i$ respectively, along with the metric perturbations, leading to $i$ derivatives of the perturbation fields in the equations of motion. We also include the Lagrangian ${\cal L}_{\alpha^0\alpha}^i$ involving interactions between $\alpha^0$ and $\alpha_i$. 

The Lagrangians $ {\cal L}_T^i$ are given in Section \ref{sec:GR}, while ${\cal L}_{\alpha^0}^i$ are the same as ${\cal L}_{\chi}^i$, for $i=(0,1,2)$, given in Section \ref{sec:ST}, but with $\chi \rightarrow \alpha^0$. For ${\cal L}_{\alpha}^i$, we have that:
\begin{equation}
{\cal L}_{\alpha}^0= a^3 \left[T_{ \alpha S}  \alpha_i  \delta N^i  +\frac{1}{2}T_{\alpha\alpha }\alpha_i \alpha^i  \right],
\end{equation}
\begin{align}
{\cal L}_{\alpha}^1& =a^3\left[T_{\partial \alpha h+ }\delta h\partial^i \alpha_i + 2L_{\partial \alpha h\times}\delta h_{ij}\partial^i\alpha^j  + T_{\dot{\alpha}S}\delta N^i \dot{\alpha}_i  + T_{\alpha\partial N }\partial^i\delta N \alpha_i \right],
\end{align}
\begin{align}
{\cal L}_{\alpha}^2&=a^3\left[ T_{\partial\dot{\alpha}h+ }\delta h\partial^i  \dot{\alpha}_i + 2L_{\partial\dot{\alpha}h\times}\delta h_{ij}\bar{h}^{jk} \partial^i \dot{\alpha}_k + T_{\partial\dot{\alpha}N } \delta N\partial^i \dot{\alpha}_i + \frac{1}{2}L_{\dot{\alpha}\dot{\alpha}}\dot{\alpha}_i\dot{\alpha}_j\bar{h}^{ij} \right. \nonumber\\
 &  + L_{ \dot{\alpha} \dot{S}} \delta \dot{N}^i  \dot{\alpha}_i   + L_{\partial \alpha K+} \delta K\partial^i \alpha_i + 2L_{\partial\alpha K\times}\delta K^i_j\partial^j \alpha_i + T_{\partial \alpha \partial S }\partial_i \delta N^i \partial^j \alpha_j  \nonumber\\
&\left. +T_{\partial\alpha\partial\alpha+}\partial^i\alpha_i\partial^j\alpha_j  +L_{\partial \alpha\partial \alpha \times }\partial^i\alpha^j\partial_i\alpha_j \right].
\end{align}
We finally add the following interaction terms to the total gravitational action:
\begin{align}
{\cal L}_{\alpha^0\alpha}^1&=a^3 T_{\alpha\partial\alpha^0}\alpha^i\partial_i\alpha^0,\\
{\cal L}_{\alpha^0\alpha}^2&=a^3 T_{\alpha\partial\dot{\alpha}^0}\alpha^i\partial_i\dot{\alpha}^0 ,
\end{align}
and ${\cal L}_{\alpha^0\alpha}^0=0$. As in the previous sections, we have integrated by parts and grouped coefficients to simplify notation. In Appendix \ref{App:CoeffVector} we give the dictionary for the coefficients $T_*$ in terms of the $L_*$ for the Lagrangians ${\cal L}_\alpha^i$ and ${\cal L}_{\alpha^0\alpha}^i$. 
Since we will also be coupling a matter scalar field $\varphi$, we must include the quadratic matter action given in eq.~(\ref{SecondSmv2}) in the total quadratic action.

Moving on to Step 3, we write the total quadratic action $S_G^{(2)}+ S_M^{(2)}$ in terms of the scalar-type perturbation fields ($\Phi$, $B$, $\Psi$, $E$, $\alpha^0$, $\alpha$ and $\delta \varphi$), find the corresponding Noether identities and solve the associated Noether constraints. After solving the Noether constraints we find that the total quadratic action depends on the following 10 free coefficients: 
\begin{align}
M^2&= 2L_{KK\times},\label{DefM2} \\
\alpha_{D1}&=\frac{L_{\dot{N}\dot{N}}}{M^2}=\frac{A^2}{M^2}L_{\dot{\alpha}^0\dot{\alpha}^0},\label{DefAD1}\\
\alpha_{D2}&= -2\frac{L_{K\dot{N}}}{M^2}, \\
\alpha_{D3}&=-\alpha_{D2}+ 2\frac{T_{N\partial \dot{S} }}{M^2}, \\
\alpha_T&= \frac{2}{M^2}\left(L_R+4L_{Rh\times}\right)-1,\label{AlphaTVT}\\
\alpha_H&=\frac{2}{M^2}\left(T_{NR}+2L_{h\partial^2N\times}\right)-1,\\
\alpha_{V0} &= \frac{1}{2HM^2}\left(T_{NK} -3HL_{KK+}\right)-\frac{3}{2},\\
 \alpha_{V1}&=\frac{1}{M^2}\left(T_{\partial S\partial S\times}+T_{\partial S\partial S+}-4L_{K\partial S+}-4T_{\partial \dot{S}h+}-L_{KK\times}\right)+1,\\
\alpha_{V2}&=\frac{T_{SS}}{2H^2M^2}, \\
\alpha_{V3}& =\frac{1}{M^2}L_{\dot{S}\dot{S}}=\frac{A^2}{M^2}T_{\dot{\alpha}\dot{\alpha}}.\label{DefAV3}
\end{align}
Note that we also have three unknown background functions $a$, $A$ and $\varphi_0$, but one relation between $a$ and $\varphi_0$ given by eq.~(\ref{N4}). Thus, the linear cosmological evolution of the most general linearly diffeomorphism-invariant vector-tensor theory is parametrized by a total of twelve free functions of time. Note that in solving the Noether constraints we assumed $\dot{A}\not=0$, and therefore for cases with constant $A$ the free functions might change.

Since the final quadratic gravitational action is somewhat unwieldy, we do not show it explicitly in this paper. Instead, we highlight some interesting aspects of its form. The first three coefficients $\alpha_{Di}$ appear in the action multiplying time derivatives of $\Phi$. From eq.~(\ref{DefAD1}) we can see that the presence of the dynamical terms for $\Phi$ ($\delta N$) are tightly related to the presence of those for $\alpha^0$, as $L_{\dot{N}\dot{N}}$ (the coefficient of $\dot{\Phi}^2$ as seen in eq.~(\ref{LT2})) is proportional to $L_{\dot{\alpha}^0\dot{\alpha}^0}$ (the coefficient of $(\dot{\alpha}^{0})^2$ as seen in eq.~(\ref{Lchi2})). The same happens for $B$ and $\alpha$, as can be seen in eq.~(\ref{DefAV3}). All the terms we have mentioned are not present in scalar-tensor theories, as $B$ and $\Phi$ are auxiliary variables in such cases. Furthermore, even if we eliminate all the terms leading to time derivatives of $B$ and $\Phi$, i.e.~set $\alpha_{D1}=\alpha_{D2}=\alpha_{D3}=\alpha_{V3}=0$, the remaining gravitational action still has different quadratic metric interaction terms, compared to the ones in eq.~(\ref{SFourthOrderFinal}), namely $(\partial B)^2$, $(\partial\Phi)^2$, and $(\partial^2\dot{E})^2$. Note also that two of the ten coefficients in the final action are the same as those present for a scalar-tensor theory: $\alpha_T$ and $\alpha_H$. However, they do not enter the action in exactly the same way; for instance, both $\alpha_T$ and $\alpha_H$ multiply a term of the form $(\partial\Phi)^2$. 

A detailed analysis of the physical propagating DoFs and the stability of this class of theories is beyond the scope of this paper. However, we comment on the fact that if all the free coefficients are nonzero, we might naively think that this gravitational action propagates four physical scalar DoFs, as $E$, $\Psi$, $\Phi$, $B$, $\alpha$ and $\alpha^0$ are dynamical fields (and as there are two scalar gauge parameters). This would suggest the propagation of unstable modes, given that a well behaved vector-tensor (Lorentz-invariant) theory is expected to propagate at most two scalar DoFs: the helicity-0 modes from the massive spin-1 and spin-2 particles. For this reason it is instructive to make the following redefinition of the vector perturbation fields:
\begin{equation}
\tilde{\alpha}=\alpha + A B, \quad \tilde{\alpha}^0=\alpha^0+A\Phi.
\end{equation}
With this redefinition the fields $B$ and $\Phi$ become auxiliary variables in the action (i.e.~do not have any time derivatives), while $\tilde{\alpha}$ and $\tilde{\alpha}^0$ are dynamical. In this way, it is clear that the action will propagate at most two scalar DoFs. Furthermore, well-known linearly diffeomorphism-invariant vector-tensor theories propagate only one healthy scalar DoF. In this context, we notice that extra conditions on the coefficients might reduce the number of physical DoFs to one. For instance, if we set $\alpha_{V3}=0$, $\tilde{\alpha}$ becomes an auxiliary variable, or, alternatively, if we set $\alpha_{D1}=0$ then $\tilde{\alpha}^0$ becomes an auxiliary variable. Such cases should encompass the generalizations of the Proca action studied in \cite{Heisenberg:2014rta,Allys:2015sht,Jimenez:2016isa}. As we will see in the next subsection, there are alternative ways of constructing vector-tensor theories propagating only one physical scalar DoF, by incorporating extra constraints. 

It is interesting to see what happens if we also impose a $U(1)$ gauge symmetry on the vector field. In this case the quadratic action is invariant under
\begin{eqnarray}\label{GaugeMaxwell2}
\alpha^\mu\rightarrow \alpha^\mu+\partial^\mu \varepsilon,
\end{eqnarray}
where $ \varepsilon $ is an arbitrary infinitesimal parameter, independent of the other two scalar gauge parameters in the linear coordinate transformation. After solving the Noether constraints associated to the U(1) gauge symmetry we find that 
\begin{equation}
\alpha_{D1}=\alpha_{D2}=\alpha_{H}=\alpha_{V0}=\alpha_{V2}=0, \quad \alpha_{V1}=1,\quad \alpha_{T}=\frac{d \ln M^2}{d \ln a}, \quad \alpha_{D3}=-4\alpha_{V3},
\end{equation}
along with eq.~(\ref{M2eq}) and the final quadratic action depends on two free coefficients $M$ and $\alpha_{V3}$. In addition, we have only one free function describing the background $A$, as $a$ is related to $M$ through eq.~(\ref{M2eq}). In general, this action does not propagate any physical scalar DoF, because it has three dynamical fields $E$, $\Psi$ and $\tilde{\alpha}$ and three gauge parameters inducing redundancies rendering these fields unphysical. The Einstein-Maxwell theory is one example of this case. 

\subsection{Einstein-Aether theory}

As mentioned above, there are different ways of constructing a gravitational action with a vector and tensor field that propagates only one scalar DoF. Here we show one special case in which we introduce an additional constraint. Specifically, we will add the Einstein-Aether constraint: 
\begin{equation}\label{ScFull}
S_{c}=\int d^4x \; \sqrt{-g}\lambda \left(A^\mu A_\mu +1\right),
\end{equation}
to the gravitational action $S_G$, where $\lambda$ is a Lagrange multiplier, and gives the constraint $A^\mu A_\mu=-1$. In particular, $\lambda$ is an extra scalar field whose perturbation transforms under linear coordinate transformations in the same way as $\delta\chi$ in eq.~(\ref{gauge:scalar}). The presence of the new field $\lambda$ imposes an extra background equation of motion: $A=1$, while all the other background equations are the same as those in the general vector-tensor case presented previously, where now all the coefficients $L_*$ are functional derivatives of the total gravitational action (which now includes $S_c$). 

The Einstein-Aether constraint will contribute the following second-order terms to the total action:
\begin{align}
S_c^{(2)}&=\int d^4x \; a^3\left\{ -2 \delta \lambda\left(\alpha^0 + \delta N\right)\right\}, \label{Sc2}
\end{align}
where we have expanded the Lagrange multiplier as $\lambda=\lambda_{0}+\delta \lambda$. Notice that the second-order Taylor expansion of eq.~(\ref{ScFull}) will also lead to quadratic terms in the metric perturbations, but we do not show them in eq.~(\ref{Sc2}) as they are taken into account in $\mathcal{L}_T^i$ (i.e.~any quadratic metric term from $S_c$ contributes to the action via changing the explicit from of the coefficients $L_*$ and $T_*$ in the metric Lagrangians). In addition, from eq.~(\ref{Sc2}) we note that the equation of motion for $\delta \lambda$ gives the Lagrange constraint $\delta N+\alpha^0=0$. As expected, this constraint corresponds to the linear expansion of the full constraint $A^\mu A_\mu =-1$.

As in the general vector-tensor case, we follow Step 3 to express the total quadratic action $S_{G}^{(2)}+S_M^{(2)}+S_c^{(2)}$ in terms of the scalar perturbations, and impose that it is invariant under linear infinitesimal gauge transformations. After solving the Noether constraints we find an action depending on ten free coefficients (of which three are different to those present in the general vector-tensor case). However, after solving the Lagrange constraint $\alpha^0=-\delta N$ the dependence on some coefficients vanishes, while the rest combine in such a way that the final quadratic gravitational action depends on four coefficients only. The final action is:

\begin{align}\label{EAaction}
S_G^{(2)}&=\int d^4x\;  a^3 \left\{ M^2 \left[ \alpha_{V3}\frac{1}{2}\partial_i\dot{\hat{\alpha}}\partial^i\dot{\hat{\alpha}} - \alpha_{V3}\hat{\Phi}\partial^2\dot{\hat{\alpha}} +\frac{1}{2}\alpha_{V3}\partial_i\hat{\Phi}\partial^i\hat{\Phi}  -\frac{2}{3}\partial^2\hat{B}\partial^2\hat{\alpha}\right.\right. \nonumber\\
& \left. +\frac{1}{3}\partial^2\hat{B}\partial^2\hat{B}  +\frac{1}{3}\partial^2\hat{\alpha}\partial^2\hat{\alpha} + H^2 \alpha_{V5}\partial_i\hat{\alpha}\partial^i\hat{\alpha} + H\alpha_{V4}\partial_i\hat{\alpha}\partial^i\hat{\Phi}\right]+ \frac{(\rho_m+P_m)}{\dot{H}H}\left[\frac{3}{2}H^3\hat{\Phi}^2 \right. \nonumber\\
&  + H^2\hat{\Phi}\partial^2 \hat{B}+\frac{1}{6}H\partial^2\hat{\alpha}\partial^2\hat{\alpha} -\frac{1}{3}H\partial^2\hat{\alpha}\partial^2\hat{B}+\frac{1}{6}H\partial^2\hat{B}\partial^2\hat{B}  + \dot{H}\partial_i\Psi\partial^i\hat{B}-3\dot{H}H\Psi\hat{\Phi}\nonumber\\
& \left. +\frac{\dot{H}}{2H} \partial_i\Psi \partial^i\Psi- \frac{3\dot{H}^2}{2H}\Psi^2\right]  -  P_m\left( \frac{3}{2}\Psi^2-a^2\Psi\partial^2 E -\frac{a^4}{2}\partial^2 E\partial^2E \right)\nonumber\\
& \left. -\rho_m\left(\frac{1}{2}\Phi^2+\frac{1}{2} B\partial^2 B +3\Phi \Psi- a^2 \Phi \partial^2 E\right) \right\},
\end{align}
where, to simplify our expression, we have defined:
\begin{align}
\hat{\alpha}&= \tilde{\alpha}  -\frac{\Psi}{H}  ,\\
\hat{B}&=B -\frac{\Psi}{H} - a^2\dot{E},\\
\hat{\Phi}&=\Phi + \frac{\dot{\Psi}}{H} -  \Psi\frac{\dot{H}}{H^2} ,
\end{align}
and where $\tilde{\alpha}=\alpha+B$. Note that in eq.~(\ref{EAaction}) the last two sets of parentheses will cancel with their corresponding counterparts from the matter action given in eq.~(\ref{SecondSmv2}). Also, note that $S_c^{(2)}=0$, as we have solved the Lagrange constraint. 
In the final quadratic action given by eq.~(\ref{EAaction}) $M^2$ and $\alpha_{V3}$ are given by eq.~(\ref{DefM2}) and (\ref{DefAV3}) respectively, while the other two free coefficients are given by:
\begin{align}
\alpha_{V4}& =\frac{1}{M^2H}\left[ -2\dot{L}_{K\partial S\times} + 3\left(L_{KK+}-2L_{K\partial S\times}\right)H-4L_{hK\times}+4L_{h\partial S\times}\right]+\frac{d\ln M^2}{d\ln a} \nonumber\\
&  - \alpha_T+ 3, \nonumber\\
\alpha_{V5}& =\frac{3}{2M^2H}\left( \dot{L}_{KK+}+HL_{KK+}\right)-\frac{1}{2}\left(\alpha_{V4}-1\right) \frac{d\ln M^2}{d\ln a}-\frac{\dot{\alpha}_{V4}}{2H}-\frac{\alpha_{V4}\dot{H}}{2H^2}+\alpha_T\nonumber\\
& -\frac{\alpha_{V4}}{2}+\frac{3}{2}, 
\end{align}
where $\alpha_T$ is given by eq.~(\ref{AlphaTVT}). This final quadratic action encompass all vector-tensor theories that include the Einstein-Aether constraint in eq.~(\ref{ScFull}).

From eq.~(\ref{EAaction}) we can see that when solving the Lagrange constraint, $\Phi$ becomes an auxiliary variable. Thus the final action has two auxiliary fields, $B$ and $\Phi$, and three dynamical fields $E$, $\Psi$ and $\tilde{\alpha}$, with no dependence on $\delta \lambda$ and $\alpha^0$. This action is still gauge invariant under linear infinitesimal coordinate transformations; the Lagrange constraint does not fix any preferred gauge because $\Phi+\alpha^0$ is a gauge-invariant quantity. Therefore, this action propagates only one physical scalar DoF, as the two scalar gauge parameters render two dynamical fields unphysical.

The final action in eq.~(\ref{EAaction}) depends explicitly on four coefficients, while the background has only one free function $a$. Therefore this cosmological model is parametrized by five free functions in total. Notice that we expect $\lambda_{0}$ to appear in the background equations of motion, but we do not count it as an extra free function, since it can be eliminated by appropriately combining the background equations. Thus $\lambda_{0}$ is not directly observable. In addition, $A$ and $\varphi_0$ do not count as free parameters either because $A$ is fixed to be $A=1$, and $\varphi_0$ will be related to $a$ by the matter background eq.~(\ref{N4}).

\section{Discussion}
\label{sec:discussion}

In this paper we have constructed a method for parametrizing the most general, local, quadratic actions for linear cosmological perturbations. This is a crucial step towards identifying how many free functions fully characterize the landscape of gravitational theories in the linear cosmological regime. Our systematic method for finding such actions, given a field content and (set of) gauge symmetries, consists of the following three main steps: 
\begin{itemize}
\item[1.] Assume a given number and type(s) of fields present in the theory (gravity and matter). Given an ansatz for the cosmological background, consider linear perturbations around that background for each field. Finally, choose what gauge symmetries to impose on the quadratic action determining the evolution of these perturbations.

\item[2.] Construct the most general local quadratic gravitational action, given the content field set in Step 1. Start with an unperturbed fundamental gravitational action $S_G$, a functional of a set of building blocks $\vec{\Theta}$ containing all the fields and their derivatives (up to some truncating maximum order). Find the perturbed set of building blocks $\delta \vec{\Theta}$, given the linear perturbations of the fields, and Taylor expand $S_G$ up to second order in $\delta \vec{\Theta}$. Finally, add some known matter action $S_M$ and Taylor expand in the same way. The first-order total action $S_G^{(1)}+ S_M^{(1)}$ leads to the background equations of motion, while the second-order total action $S_G^{(2)}+S_M^{(2)}$ determines the evolution of the linear cosmological perturbations. The form of $S_G^{(2)}$ should be that of an action including all possible covariant quadratic interactions between the linear perturbation fields. Each term in this action has an \textit{a priori} free coefficient in front, which is a functional derivative of the fundamental action $S_G$ evaluated at the background.

\item[3.] Find the most general linearly gauge-invariant quadratic action for perturbations. Consider $S_G^{(2)}+ S_M^{(2)}$ from Step 2, and find the Noether identities associated to the desired gauge symmetry. Each gauge parameter will lead to a Noether identity, which in turn will lead to a number of Noether constraints which are, in general, linear ordinary differential equations of the free coefficients in $S_G^{(2)}$. After solving the system of Noether constraints and replacing the results in $S_G^{(2)}$, we obtain the most general quadratic gravitational action for linear cosmological perturbations for that particular field content and set of symmetries. From this result it is straightforward to identify the number of free parameters describing the linear cosmological evolution of the universe, and the number of physical DoFs propagating.
\end{itemize}

In this procedure, the free parameters characterizing the quadratic action for perturbations are related to properties of fundamental gravitational theories. This makes the procedure useful for translating cosmological constraints into constraints on fundamental actions, as well as for straightforwardly finding where a given gravity theory lies in the space of these free parameters. In addition, since our method is very systematic, all the calculations presented in this paper are easily generalizable to different backgrounds, to include extra gravitational fields, and different gauge symmetries. 

In this paper we have applied the above procedure to a few different cases, summarized in Table \ref{SummaryTable}. 
\begin{table}[h!]
\centering
\begin{tabular}{| c || c || c || c || c |}
\hline
Fields & Der. & Free Functions & ST DoFs & Theories \\ \hline \hline
 $g_{\mu\nu}$ & 2 & $M$ & 0 & GR\\ \hline  
 $g_{\mu\nu}$, $\chi$  & 2 & $M$, $\alpha_{\{K, T, B\}}+2$ & 1 & Horndeski \\ \hline
$g_{\mu\nu}$, $\chi$  & 3 & $M$, $\alpha_{\{K, T, B, H\}}+2$ & 1 & Beyond Horndeski\\ \hline
\rowcolor[gray]{.95}
$g_{\mu\nu}$, $\chi$  & 4 & $M$, $\alpha_{\{K, T, B, H, P\}}$, $\alpha_{Q_{\{1,2,3,4,5\}}}+2$ & 1 & 4\textsuperscript{th} Scalar-Tensor\\ \hline
\rowcolor[gray]{.95}
$g_{\mu\nu}$, $A^{\mu}$  & 2 & $M$, $\alpha_{\{ T, H\}}$, $\alpha_{D_{\{1,2,3\}}}$, $\alpha_{V_{\{0, 1,2,3\}}}+2$ & 2 & 2\textsuperscript{nd} Vector-Tensor\\ \hline
\rowcolor[gray]{.95}
$g_{\mu\nu}$, $A^{\mu}$, $\lambda$  & 2 & $M$, $\alpha_{V_{\{3,4,5\}}}+1$ & 1 & Einstein-Aether\\ \hline
\end{tabular}
\caption{\label{SummaryTable} In this table we summarize the results found throughout the paper for cases in which invariance under linear coordinate transformations was assumed. The first column indicates the field content of the gravitational action. In all cases we also added a matter scalar field $\varphi$ whose presence is omitted in this table. The second column indicates the maximum number of derivatives of the perturbation fields allowed in the equations of motion. Note that in the cases where this number is higher than 2, we assumed a maximum of two time derivatives, but allowed higher spatial derivatives. The third column shows the free coefficients parametrizing the quadratic action, while the $+1$ or $+2$ counts the number of extra free background functions. The fourth column shows the maximum number of scalar-type DoFs propagated by the quadratic gravitational action. In all cases the complete quadratic theory would propagate one more matter scalar DoF. The fifth column shows theories that are encompassed by the corresponding parametrization. The three grey rows show new parametrizations of fourth-order derivative scalar-tensor theories and second-order derivative vector-tensor theories, including Einstein-Aether.}
\end{table}
We have first applied the procedure to a purely metric theory, leading to second-order derivatives in the equations of motion. In this case we found one free coefficient $M$, a function of time, describing the cosmological background and linear evolution of the universe. We also found that these quadratic gravitational actions do not propagate any scalar-type DoF. When $M=M_{Pl}$ we recover GR. We do not uniquely obtain the quadratic action for GR in this case, as GR is fully diffeomorphism-invariant, but we only required linear diffeomorphism invariance. In other words, there is more to a full theory than its quadratic action, and there are additional constraints not captured by any formalism based on linearized perturbations. Therefore, the fundamental theories described by the parameter $M$ could break the full diffeomorphism invariance, or maybe propagate extra DoFs that are only present at higher perturbative order. This also means that, in general, not all the possible values of the free parameters will be associated to healthy fundamental theories. This first case then highlights the fact that even with an accurate measurement of the free parameters, we will never be able to completely pin down the landscape of fundamental theories to only one by using linear cosmological perturbation theory alone.

We also applied our procedure to scalar-tensor gravity theories, leading to second, third and fourth-order derivatives in the equations of motion. The first two cases are well known, and the quadratic actions found encompass the theories of Horndeski and Beyond Horndeski. We also analyzed the fourth-derivative case and identify a total of 13 free functions of time, describing the background (2) and linear (11) cosmological evolution of the universe, of which 6 are new compared to the third-derivative case. In all these cases the quadratic gravitational action propagates only one scalar-type DoF. The procedure could also be applied systematically to allow higher-order derivatives, and we would most likely generate more free parameters encompassing even more theories. 

Finally, we applied the procedure to vector-tensor theories, leading to second-order derivatives in the equations of motion. We found a total of 12 free functions of time describing the background (2) and linear (10) cosmological evolution of the universe. In general, these quadratic gravitational actions propagate two scalar-type DoFs, although there could be only one when some specific parameters are zero. As an alternative case of a vector-tensor theory propagating only one scalar-type DoF, we applied the procedure to theories of gravity with an Einstein-Aether constraint. We found a total of only 5 free parameters of time describing the background (1) and linear perturbations (4). 

In all the cases presented in this paper we minimally coupled the metric field to a matter scalar field; however, the same results hold for a general perfect fluid. In addition, we only analyzed scalar-type perturbations, but the same free parameters will also describe vector and tensor-type linear perturbations, around homogeneous and isotropic backgrounds. The specific form of the quadratic action of vector and tensor perturbations is left for future work. 

We remark that the field content, and more specifically how all fields transform under a given gauge symmetry, is crucial in determining the final form of the quadratic action. For instance, we could apply the same procedure to a gravitational theory with a tensor coupled to a generalized scalar-type field $\chi$, whose linear perturbation $\delta \chi$ transforms under linear coordinate transformations as: 
\begin{eqnarray}
\delta \chi\rightarrow \delta \chi+G_0\pi+G_1\dot{\pi}+G_2\epsilon+G_3\dot{\epsilon},
\end{eqnarray}
where $G_*$ are unknown functions of the background and $\pi$, $\epsilon$ are arbitrary functions defined in Appendix \ref{app:gauge}. After applying the three steps above, focusing on scalar-type perturbations, and allowing up to two derivatives of the fields, we could get very different results to those of scalar-tensor theories. If $G_i\not=0$ for $i=(0,1,2,3)$ the final quadratic gravitational action reduces to that of generalized GR found in Section \ref{sec:GR}, and thus no scalar-type DoFs are propagated by the resulting action. On the contrary, if $G_2=G_3=0$, and $G_0=\dot{G}_1$, i.e.~when $\delta \chi$ is the linear time-like scalar-type component of a perturbed vector field $A^\mu$, the final gravitational quadratic action propagates a maximum of two scalar-type DoFs. $\delta\chi$, $\Phi$, $E$ and $\Psi$ are dynamical fields in the quadratic action, but there are two scalar gauge parameters inducing two non-physical fields. As in scalar-tensor theories, in this case we introduced only one extra field $\chi$ to the gravitational theory, but the resulting number of propagating DoFs is different. We see then the importance of our Step 1 in determining the space of gravitational theories under consideration, by defining the gauge transformation properties of the extra degrees of freedom. 

The ultimate goal is to construct an action that spans as large a swathe of the landscape of gravitational theories as possible. To do so, 
in future we will tackle theories with two tensor fields (or metrics), and thus parametrize current bigravity theories. We hope to bring these results together with those of the current paper, and propose a completely general parametrization for theories of gravity with one propagating scalar-type degree of freedom. This extends the widely-used parametrization of \cite{Bellini:2014fua} that arises in Horndeski theories, and is a substantial step towards achieving a general parametrization which transcends scalar-tensor theories. It will also allow us to identify the subspace of effective parameters in the PPF approach \cite{Baker:2012zs} which is, at the moment, still the most general parametrization of gravitational theories currently available. We will then analyze the quasi-static limit of our new formalism, map out the region of stability of these theories, and ultimately develop a numerical tool which can be used for analyzing data from forthcoming large-scale structure surveys such as Euclid, SKA, LSST and WFIRST.

\begin{acknowledgments}

PGF is extremely grateful to Jerome Gleyzes for his time and patience in explaining many conceptual and technical aspects of the EFT approach. We thank D. Alonso, E. Bellini, A. Buonanno, L. Pogosian, M. Raveri, A. Silvestri, F. Vernizzi, H. Winther and M. Zumacalerragui for useful discussions. ML was funded by Becas Chile. PGF acknowledges support from STFC, BIPAC and a Higgs visiting fellowship from the University of Edinburgh and acknowledges hospitality of the Albert Einstein Insitute in Berlin during which this project was initiated. JN acknowledges support from the Royal Commission for the Exhibition of 1851, BIPAC and Queen's College, Oxford. TB is supported by All Souls College, Oxford, and the US-UK Fulbright Commission. The {\it xAct} package for Mathematica \cite{xAct} was used in the computation and check of some of the results presented here. 

\end{acknowledgments}

\appendix
\section{3+1 decomposition}
\label{app:ADM}
In this section we present the 3+1 decomposition of the metric used throughout the paper. The space-time metric $g_{\mu\nu}$
can be decomposed as follows:
\begin{eqnarray}
g_{\mu\nu}=-n_\mu n_\nu+h_{\mu\nu},
\end{eqnarray}
where $n^\mu$ is a time-like unit vector satisfying $n^\mu n^\nu g_{\mu\nu}=-1$. Note that this means that $h_{\mu\nu}n^\nu=0$, and then $h_{\mu\nu}$ describes 3-dimensional space-like hypersurfaces normal to $n^\mu$. 

If we define the {\it lapse} and {\it shift} functions through:
\begin{eqnarray}
n^0&=&\frac{1}{N}, \\
n^i&=&-\frac{N^i}{N},
\end{eqnarray}
then the metric components become:
\begin{eqnarray}
g_{00}&=&-{N^2}+h_{ij}N^iN^j, \\ g_{0i}&=&h_{ij}N^j, \\ g_{ij}&=&h_{ij}. 
\end{eqnarray}

In this setting, we can construct the Ricci curvature for the 3-dimensional space, $R_{\mu\nu}$ in terms of $h_{ij}$ and the corresponding three dimensional covariant derivatives, as well as the extrinsic curvature $K^{\mu}_{\phantom{\mu}\nu}$:
\begin{eqnarray}
K^{\mu}_{\phantom{\mu}\nu}\equiv h_\nu^{\phantom{\nu}\rho}\nabla_\rho n^\mu,
\end{eqnarray}
which satisfies
\begin{eqnarray}
K^{\mu}_{\phantom{\mu}\nu}n^\nu=K^{\mu}_{\phantom{\mu}\nu}n_\mu=0. 
\end{eqnarray}
Specifically, in terms of the lapse and shift functions, the extrinsic curvature can be rewritten as:
\begin{eqnarray}
K_{ij}=\frac{1}{2N}\left({\dot h}_{ij}-D_iN_j-D_jN_i\right),
\end{eqnarray}
where ${\dot h}\equiv dh/dt$ and $D_i$ denotes the covariant derivatives in the 3-dimensional space described by $h_{ij}$.

For completeness, we also show the Gauss-Codazzi relation, which relates the standard 4-dimensional curvature $^{(4)}R_{\mu\nu}$ to the 3-dimensional curvature $R_{\mu\nu}$:
 \begin{eqnarray}
 ^{(4)}R=K_{\mu\nu}K^{\mu\nu}-K^2+R+2\nabla_\mu\left(Kn^\mu- n^\rho\nabla_\rho n^\mu\right),
 \end{eqnarray}
 or, alternatively, through
 \begin{eqnarray}
R_{\mu\nu}=h_\mu^\rho h_\nu^\sigma \left[ {^{(4)}}R_{\sigma\rho} +n^\alpha n^\beta {^{(4)}R}_{\rho \alpha \sigma \beta}\right]-K K_{\mu\nu}+K_{\mu \rho}K^\rho_\nu.
\end{eqnarray}

\section{Scalar perturbations}
\label{app:metricperts}
In this section we show relevant quantities in terms of the four linear scalar perturbations of the metric. 

Following the standard SVT decomposition, we consider linear perturbations around a FRW background, and write the metric components in terms of four scalar perturbation fields $B$, $\Phi$, $\Psi$ and $E$ in the following way:
\begin{eqnarray}\label{Def4Pert}
g_{00}&=& -\left(1+2\Phi\right), \nonumber \\
g_{0i}&=& \partial_iB, \nonumber \\
g_{ij}&=& a^2\left[\left(1-2\Psi\right)\delta_{ij}+2\partial_i\partial_jE\right] ,
\end{eqnarray}
where $a$ is the scale factor and depends only on the time $t$, while the four perturbations depend on time and space in general. 

Using the expressions given in Appendix \ref{app:ADM} in the 3+1 decomposition, we can express all the relevant quantities used throughout the paper in terms of the four scalar metric fluctuations:
 \begin{eqnarray}
 \delta N&=&\Phi, \\
 \delta N^i &=&\bar{h}^{ij}\partial_jB, \\
\delta h_{ij}&=&a^2\left[-2\Psi\delta_{ij}+2\partial_i\partial_jE\right], \\
 \delta_2 N &= &-\frac{1}{2}\Phi^2+\frac{1}{2}\bar{h}^{ij}\partial_iB\partial_jB, \\
 \delta \sqrt{|h|}&=&a^3\left[-3\Psi- a^2\partial^2 E\right], \\
 \delta_2 \sqrt{|h|} & =& a^3\left[ \frac{3}{2}\Psi^2 - \frac{1}{2}a^4(\partial^2 E)(\partial^2 E)-a^2\Psi \partial^2 E\right], \\
 \delta K^i_{\phantom{i}j}&=&-({\dot \Psi}+H\Phi)\delta^i_{\phantom{i}j} +
 a^2\bar{h}^{il}\partial_l\partial_j{\dot E}-\bar{h}^{il}\partial_l\partial_j B, \\
 \delta K&=& -3({\dot \Psi}+H\Phi) +
a^2 \partial^2 {\dot E}-\partial^2 B, \\
 \delta R^i_{\phantom{i}j}&=&\delta^{i}_{\phantom{i}j}\partial^2\Psi+\bar{h}^{il}\partial_l\partial_j\Psi ,\\
 \delta R&=& 4\partial^2\Psi, \\
 \delta_2 R&=& 2\left[4\Psi\partial^2\Psi-\bar{h}^{ij}(\partial_i\Psi)(\partial_j\Psi) \right]- 4a^2\partial^2\Psi\partial^2E,
 \end{eqnarray}
where $\bar{h}^{ij}=\frac{\delta^{ij}}{a^2}$ represents the background spatial metric, and also $\partial^2=\bar{h}^{ij}\partial_i\partial_j$. Here, a single $\delta$ stands for linear perturbations, while $\delta_2$ stands for quadratic perturbations, which will be needed to calculate the second-order action. 

\section{Gauge Transformations}
\label{app:gauge}
In this section we present the linear gauge transformation rules for a metric, vector and scalar field under linear coordinate transformations. 

In general, the linear transformations of any field can easily be derived from the general transformation laws. For a 2-rank tensor field $g^{\mu\nu}$, the general transformation law from a set of coordinates $x^{\mu}$ to another coordinates $\tilde{x}^{\mu}$ is given by:
\begin{equation}
\tilde{g}^{\mu\nu}(\tilde{x})=\frac{\partial \tilde{x}^\mu}{\partial x^\alpha}\frac{\partial \tilde{x}^\nu}{\partial x^\beta}g^{\alpha\beta}(x),
\end{equation}
where $\tilde{g}^{\mu\nu}$ represents the tensor in the $\tilde{x}^\mu$ coordinates. If we now consider linear transformations where $\tilde{x}^\mu=x^\mu+\epsilon^\mu$, with $\epsilon^\mu$ being an arbitrary small 4-vector, the transformation law becomes:
\begin{equation}\label{LinTrans1}
\tilde{g}^{\mu\nu}(x)\approx g^{\alpha\beta}(x) \left(\delta^{\mu}{}_{\alpha}+\partial_\alpha \epsilon^\mu \right)\left( \delta^{\nu}{}_{\beta}+\partial_\beta \epsilon^\nu\right)- \epsilon^\alpha \partial_\alpha \tilde{g}^{\mu\nu}(x),
\end{equation}
where we have Taylor expanded the coordinates of $\tilde{g}^{\mu\nu}(\tilde{x})$ up to first order in $\epsilon^\mu$. Here all partial derivatives are with respect to the $x^\mu$ coordinates. Then, up to linear order in $\epsilon^\mu$, we get
\begin{equation}\label{LinTrans2}
\tilde{g}^{\mu\nu}(x)=g^{\mu\nu}(x)+ g^{\mu\beta}\partial_\beta\epsilon^\nu + g^{\beta\nu}\partial_\beta\epsilon^\mu-\epsilon^\alpha \partial_\alpha g^{\mu\nu}(x),
\end{equation}
which can be seen as a gauge transformation where the fields change but the coordinates are kept fixed. Notice that in our last step we have used eq.~(\ref{LinTrans1}) recursively to write $\epsilon^\alpha \partial_\alpha \tilde{g}^{\mu\nu}(x)=\epsilon^\alpha \partial_\alpha g^{\mu\nu}(x) +\mathcal{O}(\epsilon^2)$. 

Now we consider linear perturbations of the metric $\delta g_{\mu\nu}$ around some background metric $\bar{g}_{\mu\nu}$, and find the gauge transformation rule for $\delta g_{\mu\nu}$ under linear coordinate transformation. From eq.~(\ref{LinTrans2}), at zeroth order the metric does not change, while at linear order we find:
\begin{equation}\label{LinTrans3}
\delta \tilde{g}^{\mu\nu} = \delta g^{\mu\nu} + {\bar g}^{\mu\beta}\partial_\beta\epsilon^\nu + {\bar g}^{\beta\nu}\partial_\beta\epsilon^\mu-\epsilon^\alpha \partial_\alpha {\bar g}^{\mu\nu},
\end{equation}
where $\delta g^{\mu\nu}$ are the linear perturbations of the inverse metric $g^{\mu\nu}$. Here it is understood that all the fields depend on the coordinates $x^\mu$.
Finally, from eq.~(\ref{LinTrans3}) we can find the transformation rule for $\delta g_{\mu\nu}$ by using that $g^{\mu\alpha}g_{\alpha\nu}=\delta^{\mu}_{\phantom{\mu}\nu}$. We find that:
\begin{equation}\label{LinTrans4}
\delta \tilde{g}_{\mu\nu} = \delta g_{\mu\nu} - {\bar g}_{\mu\beta}\partial_\nu\epsilon^\beta - {\bar g}_{\beta\nu}\partial_\mu\epsilon^\beta+\epsilon^\alpha {\bar g}_{\mu\beta}{\bar g}_{\nu\gamma} \partial_\alpha {\bar g}^{\beta\gamma}. 
\end{equation}

If we focus on the scalar-type perturbations, defined in eq.~(\ref{Def4Pert}), around a spatially-flat homogeneous and isotropic background metric, from eq.~(\ref{LinTrans4}) we find that:
 \begin{eqnarray}
  \tilde{\Phi}&=&\Phi -{\dot \pi}, \nonumber \\
  \tilde{B}&=&B+{\pi}-a^2{\dot \epsilon}, \nonumber \\
  \tilde{\Psi}&=&\Psi+ \frac{\dot a}{a}{\pi}, \nonumber \\
  \tilde{E}&=&E -\epsilon,
 \end{eqnarray}
where $a(t)$ is the scale factor. Here the dots denote derivatives with regards to the physical time $t$. Notice we have also rewritten the gauge parameter $\epsilon^\mu$ in terms of its scalar-type components as $\epsilon^\mu=(\pi,\delta^{ij}\partial_j\epsilon)$.
 
Finally, we emphasise that the same kind of analysis can be done for any type of field. For a linear perturbation of a scalar field $\chi$, expanded as $\chi=\chi_0+ \delta\chi$, the transformation under linear coordinate transformations is given by:
\begin{equation}
\delta \tilde{\chi} = \delta \chi- \epsilon^\mu \left(\partial_\mu \chi_0 \right),
\end{equation}
where $\chi_0$ is the background solution of the scalar field and $\delta \chi$ its first-order perturbation. In the case of a homogeneous and isotropic background, the transformation becomes:
\begin{equation}
\delta \tilde{\chi} = \delta \chi-\dot{\chi}_0\pi,
\end{equation}
where we have assumed that $\chi_0=\chi_0(t)$. For linear perturbations of a vector field, expanded as $A^{\mu}=A_0^\mu+\alpha^\mu$, the transformation under linear coordinate transformations is given by:
\begin{equation}
\tilde{\alpha}^\mu = \alpha^\mu +A_0^\nu \left(\partial_\nu \epsilon^\mu\right)-\epsilon^\nu\left(\partial_\nu A_0^\mu\right),
\end{equation}
where $A_0^\mu$ is the background solution of the vector field, and $\alpha^\mu$ its first-order perturbation. If we focus on scalar-type perturbations around a homogeneous and isotropic background, the transformation becomes:
\begin{eqnarray}
 \tilde{\alpha}^0 &=&  \alpha^0 + \dot{\pi}A -\dot{A}\pi, \nonumber\\
 \tilde{\alpha} &=&  \alpha +a^2 A\dot{\epsilon}, 
\end{eqnarray}
where $(A(t), \vec{0})$ is the homogeneous and isotropic background solution, and the scalar-type perturbations are such that $\alpha^\mu=(\alpha^0, \alpha^i)=(\alpha^0, \bar{h}^{ij}\partial_j\alpha)$, where $\bar{h}^{ij}$ is the 3-spatial metric from a spatially-flat FRW background metric. 

 \section{New Coefficients}
 \label{newcoefficients}
 The coefficients $T_*$ in the quadratic action for the metric are related to the original coefficients (i.e.~the functional derivatives of the gravitational Lagrangian, $L_*$) via:
 \begin{align}
 T_{hh+}&=L_{hh+}+\frac{1}{2}L_h\\
 T_{hh\times}&=L_{hh\times}+\frac{1}{2}L_h\\
 \bar{T}&=\bar{L}-3HL_K-\dot{L}_K+2L_h\\
 T_{SS}&=L_{SS}-5HL_{S\dot{S}}-\dot{L}_{S\dot{S}}\\
 T_{NN}&=2L_{N}+L_{NN}-\dot{L}_{\dot{N}}-9HL_{\dot{N}}-12HL_{\dot{N}h}-3HL_{N\dot{N}}-\dot{L}_{N\dot{N}}\\
 T_{Nh}&=\frac{1}{2}L_h+L_{Nh}-\dot{L}_{\dot{N}h}-3HL_{\dot{N}h} \\
 T_{N}&=\bar{L}+L_N-3HL_{\dot{N}}-\dot{L}_{\dot{N}}-3HL_K\\
 T_{\partial S h+}&= L_{\partial Sh+}+\frac{1}{2}L_{\partial S}\\
T_{hR+}&= L_{hR+ }+\frac{1}{2}L_R\\
 T_{NK}&= L_{NK}-L_{\dot{N}}-2L_{\dot{N}h}\\
 T_{N\partial S}&= L_{\partial S}-L_{\dot{N}}-2L_{\dot{N}h}+L_{N\partial S}-L_{\partial NS}-3HL_{\dot{N}\partial S}-\dot{L}_{\dot{N}\partial S}+3HL_{\partial\dot{N}S}+\dot{L}_{\partial\dot{N}S} \\
 T_{\partial \dot{S}h+}&= \frac{1}{2}L_{\partial \dot{S}}+L_{\partial \dot{S}h+}\\
 T_{\partial S\partial S+}&= L_{\partial S\partial S +}+ L_{\partial S\partial S \times 1} -6H\left(L_{\partial S\partial \dot{S} +}+ L_{\partial S\partial \dot{S} \times 1}\right)-2\left(\dot{L}_{\partial S\partial \dot{S} +}+ \dot{L}_{\partial S\partial \dot{S} \times 1}\right)  \nonumber\\
&- 4L_{ S\partial^2S\times 1}  +12H\left(L_{ S\partial^2\dot{S}\times 1}+L_{\dot{S}\partial^2 S\times 1}\right)+4\left(\dot{L}_{ S\partial^2\dot{S}\times 1}+ \dot{L}_{\dot{S}\partial^2 S\times 1}\right)  \\
 T_{\partial S\partial S \times}&= L_{\partial S \partial S \times 2} + 6H(L_{S\partial^2 \dot{S}\times 2}+L_{\dot{S}\partial^2 S\times 2}-L_{\partial S\partial \dot{S}\times 2})-2\dot{L}_{\partial S\partial \dot{S}\times 2} -2 L_{ S\partial^2 S\times 2} \nonumber\\
& + 2\dot{L}_{S\partial^2 \dot{S}\times 2} + 2\dot{L}_{\dot{S}\partial^2 S\times 2} \\
 T_{\partial N \partial N}&= -2L_{N\partial^2N}+H(L_{N\partial^2 \dot{N}}+L_{\dot{N}\partial^2 N}-L_{\partial N\partial \dot{N}})+\dot{L}_{N\partial^2 \dot{N}}+ \dot{L}_{\dot{N}\partial^2 N} +L_{\partial N\partial N}  \nonumber\\
& -\dot{L}_{\partial N\partial \dot{N}} -2L_{\partial^2 N} +7H L_{\partial^2\dot{N}}+\dot{L}_{\partial^2 \dot{N}} \\
 2T_{h\partial^2N+}&= L_{\partial^2 N}-HL_{\partial^2\dot{N}}-\dot{L}_{\partial^2 \dot{N}}+2 L_{h\partial^2N+}\\
T_{NR}&=L_R+ L_{NR}\\
 T_{N\partial \dot{S}}&= L_{\partial \dot{S}}+L_{N\partial \dot{S}}- L_{\dot{N} \partial S}-L_{\partial N\dot{S}}+L_{\partial \dot{N}S}\\
 T_{h\partial^3 S+ }&= L_{ h\partial^3 S+}+\frac{1}{2}L_{\partial^3 S}\\
 T_{K\partial^2 N + }&=L_{K\partial^2 N +}-L_{\partial^2\dot{N}}\\ 
 T_{\dot{N}\partial \dot{S} }&= L_{\dot{N}\partial \dot{S}}-L_{\partial \dot{N}\dot{S}}\\
T_{\partial^2 N\partial S }&= L_{\partial^3 S N}+L_{\partial^3 S} -L_{\partial^2\dot{N}}-L_{\partial N\partial^2 S}+L_{\partial^2 N\partial S}-L_{\partial^3 N S}
\end{align}

\section{Background equations}
\label{App:Background}
In this section we show the derivation of the metric background equations of motion, for a spatially-flat FRW metric. We do this by calculating the Taylor expansion of the fundamental total action (gravity and matter) up to first order on the metric perturbation fields. Let us start by finding the linear terms in the expansion of the fundamental non-perturbed gravitational Lagrangian $L_G$. From eq.~(\ref{ExpansionLgrav}) we find:
\begin{equation}
L_G^{(1)}=L_h \delta h+ L_N \delta N + L_K\delta K + L_R\delta R + L_{\dot{N}}\delta \dot{N} +L_{\partial^2 N}\partial^2\delta N+ L_{\partial S}\partial_i\delta N^i + L_{\partial \dot{S}}\partial_i\delta \dot{N}^i,
\end{equation}
where $\delta$ stands for first-order perturbations only. Thus, the linear terms of the gravitational action $S_G$ will be: 
\begin{equation}
S_G^{(1)}=\int d^4x\; \bar{L}\left(a^3\delta N+\delta \sqrt{|h|}\right) + a^3\left(L_h \delta h+ L_N \delta N + L_{\dot{N}}\delta \dot{N} +L_K\delta K\right),
\end{equation}
where $\bar{L}=L_G^{(0)}$, and we have eliminated many terms that formed a total derivative. Now we make use of the following relations:
\begin{equation}
\delta \sqrt{|h|}=\frac{1}{2}a^3\delta h; \quad \delta K = -3H\delta N +\frac{1}{2}\delta \dot{h}-\partial_i\delta N^i,
\end{equation}
to rewrite the linear action as:
\begin{align}
S_G^{(1)}&=\int d^4x\; a^3\left[ \delta N \left(\bar{L} +L_N -3HL_{\dot{N}}-\dot{L}_{\dot{N}} -3HL_K \right)  \right. \nonumber\\
&+  \left.  \frac{1}{2}\delta h \left( \bar{L} + 2L_h -3HL_K -\dot{L}_K\right)\right].
\end{align}

Now we proceed to find the linear terms from some matter action $S_M$. If we consider as matter a general perfect fluid with a stress-energy tensor $T^{\mu\nu}$, the linear expansion leads to:
\begin{equation}
S_M^{(1)}=\frac{1}{2}\int d^4x\; a^3\bar{T}^{\mu\nu}\delta g_{\mu\nu} =\int d^4x\; a^3\left(-\rho_m\delta N + \frac{P_m}{2}\delta h\right),
\end{equation}
where $\bar{T}^{\mu\nu}$ is the diagonal background stress-energy tensor for the fluid with rest-energy density $\rho_m$ and pressure $P_m$. Notice that here we have also ignored terms that formed total derivatives.

Finally, the total first-order action will be:
\begin{align}\label{LinearStotal}
S_G^{(1)}+S_M^{(1)}&= \int d^4x\; a^3\left[ \delta N \left(\bar{L} +L_N -3HL_{\dot{N}}-\dot{L}_{\dot{N}} -3HL_K-\rho_m \right) \right. \nonumber\\
&+ \left.  \frac{1}{2}\delta h \left( \bar{L} + 2L_h -3HL_K -\dot{L}_K+P_m\right)\right].
\end{align}
Now we notice that the equations of motion of the perturbation fields will have zeroth-order terms coming from the total linear action, and first-order terms coming from the total quadratic action. Since the resulting equations of motion must be satisfied order-by-order, we will have, in particular, that the total contribution from zeroth-order terms will vanish. Therefore, both brackets in eq.~(\ref{LinearStotal}) must be zero, leading to the following two metric background equations:
\begin{align}
\bar{L} +L_N -3HL_{\dot{N}}-\dot{L}_{\dot{N}} -3HL_K&=\rho_m\nonumber\\
 \bar{L} + 2L_h -3HL_K -\dot{L}_K&=-P_m.
\end{align}
Notice that the total linear action will always be zero then, given the background equations.

 \section{New Coefficients for Scalar-Tensor Action}
 \label{app:newcoeffscalar}
 The coefficients in the quadratic action of the scalar field $\chi$ are related to the original coefficients (i.e.~the
 derivatives of the lagrangian, $L_*$) via
\begin{align}
 T_{\chi \chi } &= L_{\chi\chi}-3HL_{\chi\dot{\chi}}-\dot{L}_{\chi\dot{\chi}} \\
 T_{\chi h}&= (2L_{\chi h}+L_\chi)/2 \\
 T_{\chi N }&= L_\chi +L_{\chi N}-3HL_{\chi\dot{N}}-\dot{L}_{\chi \dot{N}} \\
 T_{\dot{\chi}h}&= \frac{1}{2}L_{\dot{\chi}}+L_{\dot{\chi}h} \\
 T_{\chi\partial S}&= L_{\chi\partial S}-3HL_{\chi \partial \dot{S}}-\dot{L}_{\chi \partial \dot{S}}-L_{S\partial\chi}+3HL_{\partial \chi \dot{S}}+\dot{L}_{\partial\chi \dot{S}} \\
 T_{\dot{\chi}N }&= L_{\dot{\chi}}+L_{\dot{\chi}N}-L_{\chi\dot{N}}\\
 T_{\partial^2 \chi h+}&= \frac{1}{2}L_{\partial^2\chi}+L_{\partial^2 \chi h+ }\\
 T_{\partial\chi\partial\chi}&= L_{\partial\chi\partial\chi} -2L_{\chi\partial^2\chi}+HL_{\chi\partial^2\dot{\chi}}+\dot{L}_{\chi\partial^2\dot{\chi}} +HL_{\partial^2\chi\dot{\chi}}+\dot{L}_{\partial^2\chi\dot{\chi}} -HL_{\partial\chi \partial\dot{\chi}}-\dot{L}_{\partial\chi \partial\dot{\chi}} \\
 T_{\dot{\chi}\partial S}&=L_{\dot{\chi}\partial S}-L_{\chi\partial\dot{S}}+L_{\partial\chi\dot{S}}-L_{\partial\dot{\chi}S} \\
 T_{\partial\chi\partial N}&= -L_{\partial^2\chi} -L_{\chi\partial^2 N}+HL_{\chi\partial^2\dot{N}} + \dot{L}_{\chi\partial^2\dot{N}} +L_{\partial\chi\partial N} -HL_{\partial\chi\partial\dot{N}}-\dot{L}_{\partial\chi\partial\dot{N}}\nonumber\\
& -L_{\partial^2\chi N} +HL_{\partial^2\chi\dot{N}}+ \dot{L}_{\partial^2\chi\dot{N}} \\
 T_{\partial^2\dot{\chi}h+}&= L_{\partial^2\dot{\chi}h+}+\frac{1}{2}L_{\partial^2\dot{\chi}}\\
 T_{\partial^2\dot{\chi}N }&= L_{\partial^2\dot{\chi}}-L_{\chi\partial^2\dot{N}}+L_{\dot{\chi}\partial^2 N}+L_{\partial\chi\partial\dot{N}}-L_{\partial\dot{\chi}\partial N}-L_{\partial^2 \chi \dot{N}}+L_{\partial^2\dot{\chi}N} \\
 T_{\dot{\chi}\partial \dot{S}}&= L_{\dot{\chi}\partial \dot{S}}-L_{\partial\dot{\chi} \dot{S}} \\
 T_{\partial^2 \chi \partial S }&= L_{\chi \partial^3 S}-L_{\partial\chi \partial^2 S}+L_{\partial^2 \chi \partial S}-L_{\partial^3 \chi S }
\end{align}

\label{app:noetherGR}
\section{Fourth order action for Scalar-Tensor theory}
\label{4thorderaction}
In this section we show the most general quadratic Lagrangians of a scalar-tensor theory involving four derivatives of the perturbation fields (at most two time derivatives, though).
\begin{align}
{\cal L}_T^4&= a^3\left[ \frac{1}{2}L_{RR+} (\delta R)^2 + L_{RR\times} \delta R^i_{j}\delta R^j_{i} + L_{R\partial\dot{S}+}\delta R \partial_i \delta \dot{N}^i +L_{R\partial\dot{S}\times } \delta R^i_j\partial_i\delta \dot{N}^j \right.\nonumber\\
& + T_{\partial\dot{S}\partial\dot{S}+}\partial_i\delta \dot{N}^i\partial_j\delta \dot{N}^j  +T_{\partial\dot{S}\partial\dot{S}\times} \bar{h}_{ik}\partial^j\delta\dot{N}^k\partial_j\delta\dot{N}^i + T_{\partial^2 N\partial \dot{S}}\partial^2\delta N\partial_j\delta\dot{N}^j  \nonumber\\
&+T_{h\partial^3\dot{S}+}\delta h\partial^2\partial_i\delta\dot{N}^i +L_{h\partial^3\dot{S}\times}\delta h_{ij}\partial^2\partial^j\dot{N}^i +L_{h\partial^3\dot{S}\odot}\delta h_{ij}\partial^i\partial^j\partial_k\delta \dot{N}^k \nonumber\\
&+\frac{1}{2}T_{\partial^2N\partial^2N}\partial^2\delta N\partial^2\delta N + T_{\partial^2 S\partial^2S\times}\bar{h}_{ij}\partial^2\delta N^i\partial^2\delta N^j + T_{\partial^2 S\partial^2S+}\partial_i\partial_k\delta N^i\partial^k\partial_j\delta N^j \nonumber\\
& +T_{h\partial^4 N+}\delta h \partial^4 \delta N  +L_{h\partial^4 N\times}\delta h_{ij}\partial^i\partial^j\partial^2\delta N+ L_{K\partial^3S+}\delta K\partial^2 \partial_j\delta N^j \nonumber\\
& + L_{K\partial^3S\odot}\delta K^i_j \partial^j\partial_i \partial_k\delta N^k  + L_{K\partial^3S\times}\delta K^i_j \partial^2 \partial_i\delta N^j +L_{K\partial^2\dot{N}+}\delta K\partial^2\delta \dot{N} \nonumber\\
& \left.+ L_{K\partial^2\dot{N}\times}\delta K^i_j\partial^j\partial_i \delta \dot{N} + \frac{1}{2}L_{\partial\dot{N}\partial\dot{N}}\partial_i\delta\dot{N}\partial^i\delta\dot{N} \right],
\end{align}

\begin{align}
 {\cal L}_\chi^{4}&=a^3\left[ L_{\partial^2\chi R+}\delta R\partial^2\delta \chi +2L_{\partial^2\chi R\times}\delta R^i_j\partial^j\partial_i\delta \chi +T_{\partial^4\chi h+}\delta h \partial^4\delta \chi   +\frac{1}{2}T_{\partial^2\chi\partial^2\chi} (\partial^2\delta \chi)^2  \right. \nonumber\\
& + 4L_{\partial^4\chi h\times }\delta h_{ij}\partial^i\partial^j\partial^2\delta \chi + \frac{1}{2}T_{\partial\dot{\chi}\partial\dot{\chi} }\partial_i\delta \dot{\chi}\partial^i\delta \dot{\chi} +T_{\partial^2N\partial^2\chi}\partial^2\delta N\partial^2 \delta \chi  \nonumber\\
&\left.+ T_{\partial^2\chi\partial\dot{S}}\partial^2\delta\chi \partial_i\delta \dot{N}^i  +T_{\partial\dot{\chi}\partial\dot{N}}\partial_i\delta\dot{\chi}\partial^i\delta \dot{N} + L_{K\partial^2\dot{\chi}+}\delta K \partial^2\delta\dot{\chi} + L_{K\partial^2\dot{\chi}\times}\delta K^i_j \partial^j\partial_i\delta\dot{\chi} \right],
\end{align}
where we have made integrations by parts and grouped some coefficients $L_*$ together into new coefficients $T_*$, for simplicity.

\section{New coefficients Vector-Tensor action}\label{App:CoeffVector}
 The coefficients in the quadratic action for the vector field are related to the original coefficients (i.e.~the
 derivatives of the lagrangian, $L_*$) via:
\begin{align}
T_{ \alpha S} &=  L_{\alpha S}-3HL_{\alpha \dot{S}}-\dot{L}_{\alpha \dot{S}}   ,\\
T_{\partial \alpha h+ } &=  L_{\partial \alpha h+ } + \frac{1}{2}L_{\partial\alpha} ,\\
 T_{\alpha\alpha } &= L_{\alpha \alpha}  -HL_{\alpha \dot{\alpha}}-\dot{L}_{\alpha \dot{\alpha}} ,\\
  T_{\dot{\alpha}S}&= L_{\dot{\alpha}S} -L_{\alpha\dot{S}},\\
T_{\alpha\partial N } &=  L_{\alpha\partial N} -L_{\partial\alpha}  -HL_{\alpha\partial\dot{N}}-\dot{L}_{\alpha\partial\dot{N}} -L_{\partial\alpha N} +HL_{\partial\alpha\dot{N}}+ \dot{L}_{\partial\alpha\dot{N}}  ,\\
T_{\partial\dot{\alpha}h+ } &=  L_{\partial\dot{\alpha}h+}+\frac{1}{2}L_{\partial\dot{\alpha}},\\
  T_{\partial\dot{\alpha}N } &=  L_{\partial\dot{\alpha}}+L_{\alpha\partial\dot{N}}-L_{\dot{\alpha}\partial N}-L_{\partial \alpha \dot{N}}+L_{\partial\dot{\alpha}N} ,\\
 T_{\partial \alpha \partial S } &= L_{\partial \alpha \partial S}-L_{\alpha \partial^2 S} -L_{\partial^2 \alpha S } ,\\
 T_{\alpha\partial\alpha^0}&= L_{\alpha\partial \alpha^0} -  L_{\partial \alpha \alpha^0} ,\\
 T_{\alpha\partial\dot{\alpha}^0} &=L_{\alpha\partial\dot{\alpha}^0} - L_{\partial \alpha\dot{\alpha}^0}- H\left(L_{\dot{\alpha}\partial\alpha^0}-L_{\partial\dot{\alpha}\alpha^0}\right)-\dot{L}_{\dot{\alpha}\partial\alpha^0}+ \dot{L}_{\partial\dot{\alpha}\alpha^0}, \\
T_{\partial \alpha\partial\alpha +}&=L_{\partial \alpha\partial\alpha +}+L_{\partial \alpha\partial\alpha \times}
\end{align}

 
\bibliographystyle{apsrev4-1}
\bibliography{RefModifiedGravity}

\end{document}